\documentclass[letterpaper,11pt]{article}

\usepackage{includes/jhep/jheppub}
\usepackage{mathtools,amssymb,amsmath,amsthm,mathrsfs}
\usepackage{physics}

\usepackage{hhline}
\newcommand{\mc}[1]{\ensuremath{\mathcal{#1}}}

\usepackage{subfig}

\usepackage{graphbox,xspace,float,color}
\usepackage[table]{xcolor}

\usepackage{pgfplots}  % drawing pictures

\pgfplotsset{compat=newest}
\usepgfplotslibrary{fillbetween}

\pgfplotsset{
  every overlay node/.style={
    draw=black,fill=white,rounded corners,anchor=north west,
  },
}

\DeclareMathSymbol{\mhyphen}{\mathord}{AMSa}{"39}

\usepackage{pifont}
\usepackage{contour}
\def\le{\left}
\def\ri{\right}

\definecolor{darkpastelred}{rgb}{0.76, 0.23, 0.13}
\definecolor{ballblue}{rgb}{0.13, 0.67, 0.8}
\definecolor{azure}{rgb}{0.0, 0.5, 1.0}
\definecolor{darkspringgreen}{rgb}{0.09, 0.45, 0.27}
\definecolor{ashgrey}{rgb}{0.7, 0.75, 0.71}
\definecolor{aurometalsaurus}{rgb}{0.43, 0.5, 0.5}
\definecolor{babyblueeyes}{rgb}{0.63, 0.79, 0.95}
\definecolor{cornflowerblue}{rgb}{.53, .80, .93}  % slightly different from online val, to match matplotlib
\definecolor{royalblue}{RGB}{99,149,236}  % slightly different from online val, to match matplotlib
\definecolor{dodgerblue}{rgb}{0.12, 0.56, 1.0}
\definecolor{forestgreen(traditional)}{rgb}{0.0, 0.27, 0.13}
\definecolor{forestgreen(web)}{rgb}{0.13, 0.55, 0.13}
\definecolor{darkgoldenrod}{rgb}{0.72, 0.53, 0.04}
\definecolor{goldenrod}{rgb}{0.85, 0.65, 0.13}
\definecolor{ochre}{rgb}{0.8, 0.47, 0.13}
\definecolor{plum}{RGB}{221,160,221}

\DeclareRobustCommand{\Sec}[1]{sec.~\ref{sec:#1}}

\DeclareRobustCommand{\Fig}[1]{fig.~\ref{fig:#1}}
\DeclareRobustCommand{\Figs}[2]{figs.~\ref{fig:#1} and \ref{fig:#2}}

\DeclareRobustCommand{\Eq}[1]{eq.~(\ref{eq:#1})}

\DeclareRobustCommand{\Reff}[1]{ref.~\cite{#1}}
\DeclareRobustCommand{\Reffs}[2]{refs.~\cite{#1} and \cite{#2}}

\usepackage{marginnote}
\usepackage[normalem]{ulem}

\definecolor{samcolor}{rgb}{0.0, 0.5, 0.0} % Ao(English)

\usepackage{tcolorbox}

\newtcolorbox{sambox}[2][]{
    colback=samcolor!5!white,
    colframe=samcolor!75!black,
    colbacktitle=samcolor!10!white,
    coltitle=samcolor!70!black,
    title={#2},fonttitle=\bfseries,#1}
\newcommand{\pythia}{\texttt{Pythia 8.309}}

\newcommand{\lqcd}{\ensuremath{\Lambda_{\text{QCD}}}}
\newcommand{\rsub}{\ensuremath{r_\text{sub}}}
\newcommand{\Rjet}{\ensuremath{R_\text{jet}}}

\definecolor{firebrick}{rgb}{0.7, 0.13, 0.13}
\definecolor{magenta}{HTML}{FF00AB}
\definecolor{forestgreen}{HTML}{00B500}
\definecolor{russett}{HTML}{805B56}

\definecolor{inclusivecolor}{rgb}{0.03, 0.27, 0.49}

\newif\ifexternalize
\externalizefalse

\ifexternalize
\usetikzlibrary{external}
\usepgfplotslibrary{external}
\fi

\usetikzlibrary{decorations}
\usetikzlibrary{decorations.text}

\tikzset{
    gluon/.style={decorate,decoration={coil,amplitude=2pt, segment length=2pt,  pre length=0cm, post length=0cm}},
    photon/.style={decorate, decoration={snake, segment length=4pt, amplitude=1.8pt,  pre length=0cm, post length=0cm}}
}

\tikzstyle{calo}=[draw=blue!60!green!50!black,fill=blue!60!green!90!black!70,
                  line width=0.6,rounded corners=0.2]
\tikzstyle{ecal}=[calo,draw=red!90!green!60!black,fill=red!85!green!90!black!80]
\tikzstyle{split_cal}=[calo,draw=russett!98!green!60!black,fill=russett]
\tikzstyle{MET}=[->,red,line width=1.2,dashed]

\usepackage{tikz-feynman}
\tikzfeynmanset{compat=1.1.0}

\pgfkeys{/pgf/decoration/.cd,
         start color/.store in=\startcolor,
         start color=black,
         end color/.store in=\endcolor,
         end color=black,
         varying line width steps/.initial=100
}
\pgfdeclaredecoration{width and color change}{initial}{
 \state{initial}[width=0pt, next state=line, persistent precomputation={%
   \pgfmathparse{\pgfdecoratedpathlength/\pgfkeysvalueof{/pgf/decoration/varying line width steps}}%
   \let\increment=\pgfmathresult%
   \def\x{0}%
 }]{}
 \state{line}[width=\increment pt,   persistent postcomputation={%
   \pgfmathsetmacro{\x}{\x+\increment}
   },next state=line]{%
   \pgfmathparse{varyinglw(\x/\pgfdecoratedpathlength)}
   \pgfsetlinewidth{\pgfmathresult pt}%
   \pgfpathmoveto{\pgfpointorigin}%
   \pgfmathsetmacro{\steplength}{1.4*\increment}
   \pgfpathlineto{\pgfqpoint{\steplength pt}{0pt}}%
   \pgfmathsetmacro{\y}{varyingcolor(100*(\x/\pgfdecoratedpathlength))}
   \pgfsetstrokecolor{\endcolor!\y!\startcolor}%
   \pgfusepath{stroke}%
 }
 \state{final}{%
   \pgfmathparse{varyinglw(1)}
   \pgfsetlinewidth{\pgfmathresult pt}%
   \pgfpathmoveto{\pgfpointorigin}%
   \pgfmathsetmacro{\y}{varyingcolor(100*(\x/\pgfdecoratedpathlength))}
   \color{\endcolor!\y!\startcolor}%
   \pgfusepath{stroke}%
 }
}

\newcommand\ccr[1]{\cellcolor{red!15}{#1}}
\newcommand\ccbl[1]{\cellcolor{blue!15}{#1}}
\newcommand\ccbr[1]{\cellcolor{brown!15}{#1}}

\def\ewocsmear{94}
\def\eecsmear{-37}
\def\sdsmear{49}

\def\ewochad{-144}
\def\eechad{-1150}
\def\sdhad{18}

\def\ewocue{271}
\def\eecue{700}
\def\sdue{298}

\definecolor{wcol}{rgb}{1.0, 0.0, 0.0}
\definecolor{wwcol}{rgb}{0.93, 0.53, 0.18}
\definecolor{wwwcol}{rgb}{1.0, 0.88, 0.21}

\newcommand\ccw[1]{\cellcolor{wcol!15}{#1}}
\newcommand\ccww[1]{\cellcolor{wwcol!30}{#1}}
\newcommand\ccwww[1]{\cellcolor{wwwcol!30}{#1}}

\def\wsmear{94}
\def\wwsmear{73}
\def\wwwsmear{69}

\def\whad{-145}
\def\wwhad{-127}
\def\wwwhad{-114}

\def\wue{271}
\def\wwue{241}
\def\wwwue{230}

\title{Energy Correlators Beyond Angles}

\author[a]{Samuel Alipour-fard}
\author[b,c]{and Wouter J. Waalewijn}

\affiliation[a]{
Center for Theoretical Physics, Massachusetts Institute of Technology,
\\
77 Massachusetts Avenue, Cambridge, MA 02139, U.S.A.
}
\affiliation[b]{
Nikhef, Theory Group, Science Park 105, 1098 XG, Amsterdam, The Netherlands
}
\affiliation[d]{
Institute for Theoretical Physics Amsterdam and Delta Institute for Theoretical Physics,
\\
University of Amsterdam, Science Park 904, 1098 XH Amsterdam, The Netherlands
}

\emailAdd{samuelaf@mit.edu}
\emailAdd{w.j.waalewijn@uva.nl}

\preprint{MIT-CTP/5828}

\abstract{
Energy correlators are theoretically simple and physically intuitive observables that bridge experimental and theoretical particle physics.
They have for example enabled the most precise jet substructure determination of the strong coupling constant to date, and recent proposals suggest that they may be used to precisely determine the top quark mass with calculable, small theoretical uncertainties.
However, existing energy correlators all measure correlations in angles between particles, from which other observables such as mass must be inferred through potentially complicated procedures.
In this work, we generalize energy correlators to enable straightforward measurements of non-angular correlations, which we call Energy Weighted Observable Correlations (EWOCs).
To enforce collinear safety, EWOCs quantify correlations between \emph{subjets} rather than particles.
The subjet radius can be tuned to control both the physical scales probed by EWOCs and their sensitivity to non-perturbative physics.
We focus on the phenomenologically relevant example of the mass EWOC, which measures mass correlations between pairs of subjets, in the task of extracting mass scales from jets.
In jet substructure determinations of the mass of a hadronically-decaying \(W\) boson, we show that the mass EWOC outperforms the angle-based energy correlator, and performs comparably to the soft-drop groomed jet mass.
As a first exploration of the theoretical properties of EWOCs, we also calculate the mass EWOC on light-quark jets and compare to results obtained with \pythia{}.
}

\begin{document}
% %%%%%%%%%%%%%%%%%%%%%%%%%%%%%%%%%%%
\maketitle

%%%%%%%%%%%%%%%%%%%%%%%%%%
% Introduction:
%%%%%%%%%%%%%%%%%%%%%%%%%%
\section{Introduction}

%=======================
% Buildup
%=======================
% Big picture context

The discovery of hadronic jets in high-energy particle collisions is one of the most important milestones in human understanding of the microscopic universe, and has solidified quantum chromodynamics (QCD) as our fundamental theory of the strong nuclear force.
Jets were originally predicted \cite{Feynman:1969wa,Bjorken:1969wi,Drell:1969wb,Cabibbo:1970mh,Berman:1971xz}, and later measured \cite{Hanson:1975fe,Wiik:1979cq,Barber:1979yr,TASSO:1979zyf,PLUTO:1979dxn,JADE:1979rke,Ali:1979em,Hanson:1981em,Ali:2010tw}, as collimated sprays of hadronic radiation produced in high-energy collisions involving the strong interaction.
The quantification of the internal structure of jets, enabled by the excellent performance of the current generation of experimental detectors, has been a particularly fruitful area of study.
The study of jet substructure has led to the development of a large class of observables that characterize hadronic radiation in particle collisions, facilitating our understanding of the physical properties of the Standard Model of particle physics (SM) and beyond.

Among collider observables with the potential to probe jet substructure, the Energy-Energy Correlator (EEC) \cite{Basham:1978bw,Basham:1978zq,Basham:1979gh} is particularly noteworthy:
it is conceptually simple, insensitive to low-energy radiation, and naturally separates physics at different scales.
Physically, the EEC measures the mean correlation in the energy passing through any two detectors separated by an angle \(\Delta\), averaged over many particle collisions:
\(
    \text{EEC}(\Delta)
    \propto
    \left\langle 
        E_\text{det.\.1} \,\,\, E_\text{det.\.2} 
    \right\rangle
\).%
\footnote{
    More concretely, this means that the contribution of each particle in an event to the EEC is weighted by the particle's energy,
    see \Eq{eec_defn}.
    This energy weighting mitigates the effects of the low-energy physics of QCD that complicates the theoretical description of other jet substructure observables.
}
From the birth of the EEC to its use in recent collider studies, the simplicity of the EEC has enabled highly accurate perturbative computations \cite{Clay:1995sd,Glover:1994vz,Kramer:1996qr,DelDuca:2016csb,Gituliar:2017umx,Dixon:2018tpg,Dixon:2018qgp,Henn:2019gkr,Luo:2019nig,Gao:2020vyx,Neill:2022lqx,Lee:2023npz,Kramer:1995qh,deFlorian:2004mp,Banfi:2002vw,Tulipant:2017ybb,Moult:2018jzp,Korchemsky:2019nzm,Dixon:2019uzg,Gao:2019ojf,Luo:2019hmp,Luo:2019bmw,Moult:2019vou,Li:2020bub,Ebert:2020sfi,Li:2021txc,Duhr:2022yyp,Chen:2023zlx,Gao:2023ivm} and has facilitated measurements of the strong coupling constant \(\alpha_s\) \cite{Martin:1986uq,DELPHI:1990sof,SLD:1994yoe,ATLAS:2015yaa,ATLAS:2017qir,dEnterria:2018cye,Kardos:2018kqj,Ali:2020ksn,dEnterria:2022hzv,ATLAS:2023tgo} including the most precise jet substructure determinations of \(\alpha_s\) to date \cite{CMS:2024mlf}.
The EEC has been investigated as a probe of jets traveling through the quark-gluon plasma (QGP) produced in heavy-ion collisions \cite{Lokhtin:2004tx,Lokhtin:2006dp,Andres:2022ovj,Barata:2023zqg,Andres:2023xwr,Yang:2023dwc,Barata:2023vnl,Barata:2023bhh,Barata:2024nqo,Barata:2024ieg} and of correlations in nuclear physics \cite{Karapetyan:2019fst,Liu:2022wop,Liu:2023aqb,Kang:2023gvg,Cao:2023oef}.
The study of the EEC has also shed light on the structure of non-perturbative QCD \cite{Nason:1995np,Korchemsky:1997sy,Korchemsky:1999kt,Dokshitzer:1999sh,Chen:2020vvp,Jaarsma:2023ell,Schindler:2023cww,Lee:2024esz}
and of quantum field theory in general \cite{Richards:1983sr,Sveshnikov:1995vi,Hofman:2008ar,Hatta:2012kn,Belitsky:2013bja,Belitsky:2013ofa,Belitsky:2013xxa,Belitsky:2014zha,Korchemsky:2015ssa,Goncalves:2014ffa,Farnsworth:2015hum,Hofman:2016awc,Kravchuk:2018htv,Kologlu:2019mfz,Chang:2020qpj,Korchemsky:2021htm,Caron-Huot:2022eqs,Chen:2023wah,Chicherin:2023gxt,Chicherin:2024ifn}.
It also been put forward as a way to search for new physics in hadronic decays of electroweak bosons~\cite{Ricci:2022htc}.
Finally, while most jet substructure observables are dominated by contributions from two-particle correlations, the EEC can be naturally extended to multi-point energy correlators which can characterize correlations between an arbitrary number of particles \cite{Chang:2022ryc,Chen:2022jhb,Yan:2022cye,Komiske:2022enw,Chicherin:2024ifn}.
Recent theoretical work suggests that multi-point generalizations of the EEC may be used to extract the mass of the top quark at the LHC \cite{Procura:2022fid,Holguin:2022epo,Holguin:2023bjf,Pathak:2023tmy,Xiao:2024rol,Holguin:2024tkz}.
%

%=======================
% Introducing project:
%=======================

However, the EEC is limited to probing angular correlations, and angular correlations are not the only correlations of interest in particle collisions.
Additional, non-trivial steps are needed when using energy correlators to extract non-angular parameters of interest.
For example, in extracting the top quark mass, the conversion of angular scales to mass scales involves the transverse momentum of top quark jets;
large experimental uncertainties in the jet \(p_T\) then directly affect the resulting extraction of the top quark mass \(m_t\).%
\footnote{
    A solution to this \(p_T\) smearing was proposed by the authors of \Reff{Holguin:2023bjf}, leveraging the \(W\)-bosons that emerge during top quark decay and the well-measured mass of the \(W\)-boson.
    In particular, they define an integrated three-point energy correlator with two peaks, at angular scales \(\zeta_\text{peak} \sim m^2 / p_T^2\) with $m=m_W$ and $m=m_t$.
    The ratio of the peak locations provides a robust measure of \(m_t/m_W\) which is insensitive to uncertainties in jet \(p_T\).
    The approach we develop here does not rely on an existing, precisely determined particle mass.
}
Similarly, heavy-ion collisions and the properties of the QGP are characterized by a wide variety of correlations extending beyond the simple angular correlations probed by energy correlators \cite{Lokhtin:2004tx,Lokhtin:2006dp,Andres:2022ovj,Barata:2023zqg,Andres:2023xwr,Yang:2023dwc,Barata:2023vnl,Barata:2023bhh,Barata:2024nqo,Barata:2024ieg}.

% -----------------------------------
% EEC Visualization
% -----------------------------------
\begin{figure}
    \centering
    \hspace{9em} 
    \subfloat[]{
        \hspace{-15em}
        \includegraphics[width=.55\textwidth]{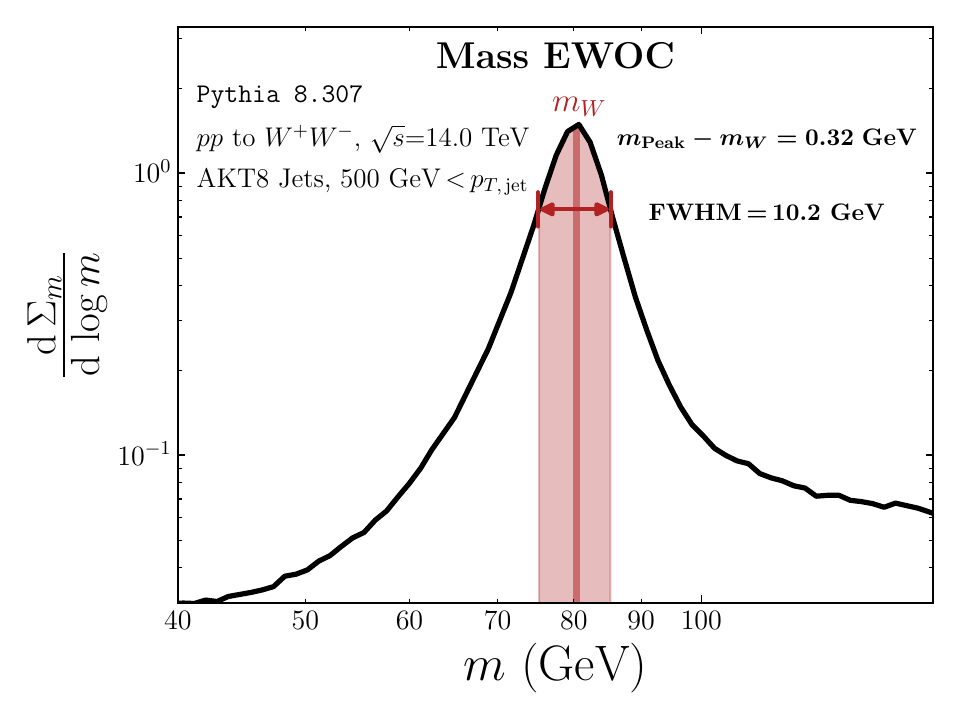}
        \hspace{-12.0em}
        \makebox[0pt][r]{% Similar to \llap
            \raisebox{5em}{%
              \includegraphics[width=.15\linewidth]{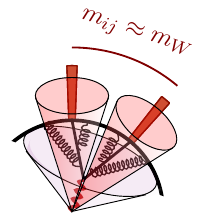}
            }
          }
    
        \label{fig:m_ewoc:pp_to_ww:with_cartoon}
    }
    \hspace{10em}
    \subfloat[]{
        \hspace{-3em}
        \includegraphics[width=.55\textwidth]{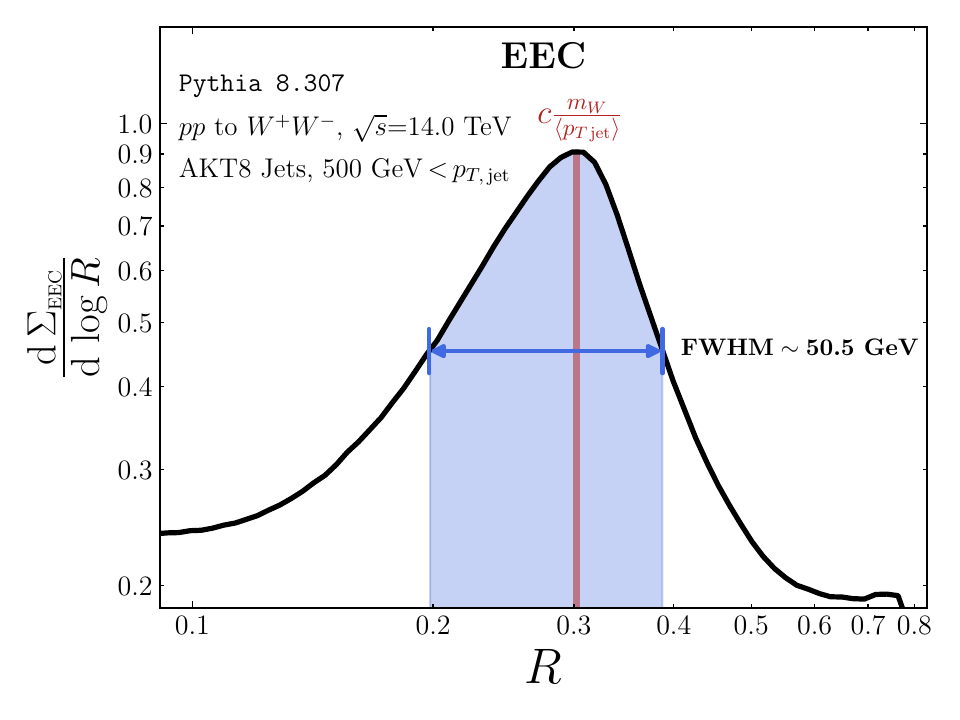}
        \hspace{-1.0em}
        \makebox[0pt][r]{% Similar to \llap
            \raisebox{9.5em}{%
              \includegraphics[width=.13\linewidth]{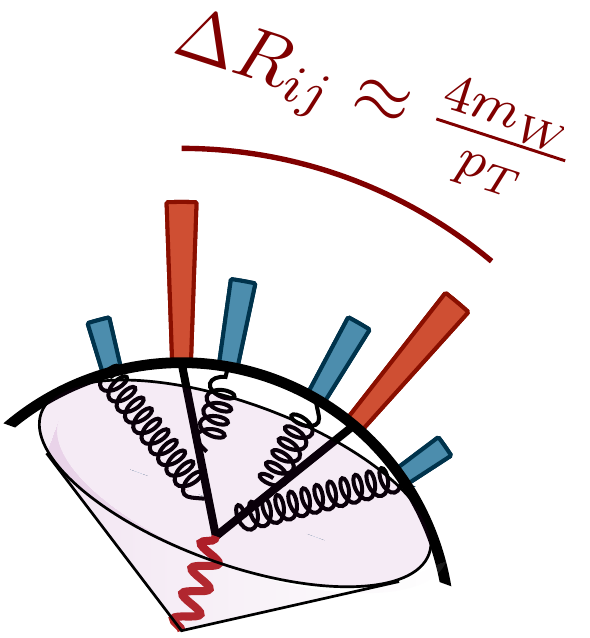}
            }
        }
        \label{fig:eec:pp_to_ww:with_cartoon}
    }
    \caption{
        Visualizations of the hadronic decay of a \(W\) boson in 14 TeV LHC events simulated with \pythia{}. 
        \hyperref[fig:m_ewoc:pp_to_ww:with_cartoon]{(a)}
        The mass EWOC, introduced in this work, which captures the mass of the of \(W\) boson directly from the pairwise mass of subjets (the red cones) of radius \(\rsub=0.3\).
        \hyperref[fig:eec:pp_to_ww:with_cartoon]{(b)}
        The EEC evaluated on pairs of particles;
        we tune the constant \(c \sim 2.5\) to extract \(m_W\) from the peak of the EEC distribution.
        The full width at half maximum (FWHM) of the mass EWOC (about 10 GeV) is significantly smaller than the mass difference associated with the FWHM of the EEC (corresponding to about 50 GeV),
        suggesting that the peak of the mass EWOC will provide a more precise estimation of \(m_W\).
    }
    \label{fig:EWOCs:visualization}
\end{figure}
% -----------------------------------

In this paper, we introduce \textbf{Energy Weighted Observable Correlations (EWOCs)} as generalizations of the EEC that preserve the strengths of energy weighting as a paradigm for extracting fundamental information from measurements, while simultaneously probing correlations in non-angular observables.
A pairwise (or 2-point) EWOC takes the form%
\footnote{
    We will present this definition in more detail in \Sec{EWOCs:EWOCs}, and a slightly more general definition with arbitrary energy weights in \Sec{EWOCs:IRC}.
    One can also generalize the pairwise EWOCs we present here to higher-point correlators that we leave to future work.
    Here, the new parametrization proposed in ref.~\cite{Alipour-fard:2024szj} would be particularly convenient to reduce computation time and enhance interpretability.
}
\begin{align}
    \label{eq:intro_ewoc_def}
    \frac{\dd \Sigma_\mathcal{O}}{\dd \chi}
    =
    \frac{1}{\sigma}
    \int \dd\sigma \,
    \sum_{
        \text{subjets }
        i,\, j
    } \,
    x_i \, x_j \,
    \delta\left(\chi \, - \, \mathcal{O}_{ij}\right)
    ,
\end{align}
where \((\dd)\sigma\) indicates the (differential) cross section for jet production in the process under consideration and the sum on \(i\) and \(j\) run over the subjets within the jet.
The energy-weighting factors \(x_{i,j}\), adopted from the EEC, ensure that higher-energy subjets contribute more to the EWOC of \Eq{intro_ewoc_def}.%
\footnote{
    For proton-proton collisions at the LHC, transverse momentum fractions \(x_i = p_{T,i}/p_{T,\text{jet}}\) are a natural choice, while energy fractions \(x_{i} = E_{i}/E_\text{jet}\) are more natural in the context of electron-positron collisions.
}
\(\mathcal{O}_{ij}\) is a user-defined  observable on pairs of subjets, such as their angular separation or invariant mass.
For example, using the two-particle angular separation in \Eq{intro_ewoc_def}, \(\mathcal O_{ij} \to \theta_{ij}\), recovers the traditional EEC of $e^+e^-$ collisions in the limit of zero subjet radius (i.e.~where the sum over pairs of subjets is replaced by a sum over pairs of particles).
However, the EWOC framework we introduce expands on the strengths of the EEC in two ways:
\begin{itemize}
    \item
        We study correlations of generic \textbf{pairwise observables}, i.e.~any observables that depend on the properties of two particles;
        this includes not only the relative angle between their momenta, probed by the EEC, but also their total invariant mass, our main focus in this work.

    \item
        We study correlations between \textbf{subjets} rather than particles within a jet;
        this isolates correlations between collective degrees of freedom (the subjets), and is essential to ensure that EWOCs are collinear-safe when considering general pairwise observables.
\end{itemize}
The radius of the subjets controls the sensitivity of EWOCs to non-perturbative effects, though it also limits the angular resolution that EWOCs can probe;
therefore, the choice of subjet radius in defining an EWOC for a particular physics context is motivated by balancing these considerations.

% -----------------------------------
% mW Estimation Table 
% -----------------------------------
\begin{table}
    \vspace{-1.0cm}

    % Table
    \begin{tabular}{|p{2.3cm}||p{2.5cm}|p{0.9cm}|p{1.9cm}|}
     \hline
     \multicolumn{4}{|c|}{
     \centering
       \textbf{Shift in \(\boldsymbol{m_W}\) Determination}
     }
     \\
     \multicolumn{4}{|c|}{
       \centering
       from the peak of each distribution
     }
     \\
     \hline
     \centering
     \vspace{5pt}
     \(\boldsymbol{\Delta}\)
     \vspace{5pt}
     &
     \ccr{
         \parbox[m]{\columnwidth}{\vspace{-5pt}\textbf{Mass EWOC}}
         \parbox[m]{\columnwidth}{\vspace{0pt}\hspace{12pt}\(k_t\) subjets,}
         \parbox[m]{\columnwidth}{\vspace{0pt}\hspace{12pt}\(r_\text{sub} = 0.3\)}
     }
     &
     \ccbl{
         \parbox[m]{\columnwidth}{\vspace{0.7cm}\textbf{EEC}}
     }
     & 
     \ccbr{
         \parbox[m]{\columnwidth}{\vspace{0.35cm}\(\boldsymbol{m}_\textbf{mMDT}\)}
         \parbox[m]{\columnwidth}{\vspace{3pt}\hspace{3pt}\(z_\text{cut} = 0.1\)}
     }
     \\
     \hline
     \hline
     \centering
     \vspace{-0.2pt}
     Smearing

     {\small{(cf \Reff{CMS:2024mlf})}}
     \vspace{10.5pt}
     &
     \vspace{8pt}
     \hspace{5pt}
     \ccr{\ewocsmear{} MeV}
     &
     \vspace{3pt}
     \ccbl{\eecsmear{} MeV}
     &
     \vspace{8pt}
     \hspace{2pt}
     \ccbr{\sdsmear{} MeV}
     \\
     \hline
     \centering
     \vspace{0pt}
     Parton vs.

     Hadron
     \vspace{10pt}
     &
     \vspace{8pt}
     \hspace{5pt}
     \ccr{\ewochad{} MeV}
     &
     \vspace{3pt}
     \ccbl{\eechad{} MeV}
     &
     \vspace{8pt}
     \hspace{2pt}
     \ccbr{\sdhad{} MeV}
     \\
     \hline
     \centering
     \vspace{0pt}
     UE (MPI)
     
     On/Off
     \vspace{10pt}
     &
     \vspace{8pt}
     \hspace{5pt}
     \ccr{\ewocue{} MeV}
     &
     \vspace{3pt}
     \ccbl{\eecue{} MeV}
     &
     \vspace{8pt}
     \hspace{2pt}
     \ccbr{\sdue{} MeV}
     \\
     \hline
    \end{tabular}
    \hspace{-20.5pt}
    \raisebox{-5.300cm}{
    \begin{tikzpicture}
    \begin{axis}[
        xbar,
        axis lines=left, 
        xlabel={\(\abs{\boldsymbol{\Delta}}\quad\)(GeV)},
        height=7.170cm,
        width=0.48\textwidth,
        xmin=0, xmax=1.16,
        ymin=0,
        ytick={0,24,48,72},
        extra y ticks={0,24,48,72},
        ymax=75,
        yticklabels={},
        extra y tick labels={}, 
        extra y tick style={grid=major},
        legend style={
         at={(0.48,1.45)},
         anchor=north,
         legend columns=1,
         /tikz/every even column/.append style={row sep=0.2cm}},
        y axis line style=-,
        ]
    \addplot [color=red!90!black, fill=red!15]
    coordinates {
        (abs{\ewocsmear}/1000,72) 
        (abs{\ewochad}/1000,48)
        (abs{\ewocue}/1000,24)
    };
    \addplot [color=blue!90!black, fill=blue!15]
    coordinates {
        (abs{\eecsmear}/1000,60) 
        (abs{\eechad}/1000,36)
        (abs{\eecue}/1000,12)
    };
    \addplot [color=brown!90!black, fill=brown!15]
    coordinates {
        (abs{\sdsmear}/1000,48) 
        (abs{\sdhad}/1000, 24)
        (abs{\sdue}/1000,0)
    };
    
    \legend{Mass EWOC, EEC, \(m_\text{mMDT}\)};

    % Units
    \iffalse
    \draw [decorate,decoration={brace,amplitude=5pt,mirror,raise=4ex}, thick]
      (axis cs: 33,50) -- (33,70)
      node[midway,xshift=1.4cm]{GeV};
    \draw [decorate,decoration={brace,amplitude=5pt,mirror,raise=4ex}, thick]
      (axis cs: 33,26) -- (33,46)
      node[midway,xshift=1.5cm,
           text width=1cm,
           align=center]{
        100s of MeV
      };
    \draw [decorate,decoration={brace,amplitude=5pt,mirror,raise=4ex}, thick]
      (axis cs: 33,2) -- (33,22)
      node[midway,xshift=1.5cm,
           text width=1cm,
           align=center]{
        100s of MeV
      };
      \fi
    \end{axis}
    
    % Adding arrow and text
    \node[anchor=west] (source) 
        at (4.9cm,5.6cm){};
    \node (destination)
        at (0.45cm,5.6cm){};
    \draw[{Round Cap}-{Latex[round]}, very thick, red!80!black](source)--(destination); 
    \node[
      above=-0.0cm of destination, 
      xshift=.5cm,
      text width=1.5cm,
      align=left
    ] {\textcolor{red!60!black}{More robust}};
    \node[
      above=0.0cm of source, 
      xshift=-.4cm,
      text width=1.5cm,
      align=right
    ] {\textcolor{red!65!blue}{Less robust}};
    \end{tikzpicture}
    }  
    \caption{
        Several measures of robustness of determinations of \(m_W\) 
        based only on the peak of the mass EWOC introduced in this work, the EEC, or the mass distribution of jets groomed using the modified mass drop tagger (mMDT).
        These are all evaluated on \pythia{} samples of a pair of hadronically-decaying \(W\) bosons at the LHC.
        While each measure of robustness may be addressed by appropriate calibration, smaller values indicate more robust determinations of \(m_W\).
    }
    \label{tab:obs_comparison}
\end{table}
% -----------------------------------

In this work, we focus on the mass EWOC, showing its power in determining the mass of a hadronically decaying particle.
In \Fig{EWOCs:visualization}, we demonstrate the use of the mass EWOC in extracting the mass of the \(W\) boson (as a proxy for a generic hadronically-decaying resonance) from simulations of \(W\) boson pair production at the LHC generated with \pythia{} \cite{Bierlich:2022pfr}. 
The EWOC framework may also be used to extract more general and phenomenologically important correlations between subjets or the masses of decaying resonances with a greater number of decay products.
For example, we expect that the three-point mass EWOC may be used as an alternative to the three-point, angle-based energy correlator for extracting the mass of top quark from its three-pronged decay \cite{Procura:2022fid,Holguin:2022epo,Holguin:2023bjf,Pathak:2023tmy,Xiao:2024rol,Holguin:2024tkz}.
While we do not explore these phenomenologically important measurements in this work, we hope that the EWOC framework will be helpful for extracting a wide variety of correlations of physical interest from future experimental measurements.

In table \ref{tab:obs_comparison}, we further evince the phenomenological value of the mass EWOC by comparing shifts in the \(m_W\) determination obtained by using the peaks of the mass EWOC, the EEC, and the mass distribution of jets groomed using the modified mass drop tagger (mMDT, which far outperforms the ungroomed jet mass) \cite{Dasgupta:2013ihk,Larkoski:2014wba}.
For each distribution, \(m_W\) is estimated by performing a quadratic fit of each distribution near the peak at \(m_W\).
In particular, we show the amount by which each estimation of \(m_W\) varies due to:
\begin{itemize}
    \item 
    an emulation of experimental detector effects via Gaussian smearing of particle momenta by 3\% for photons, 5\% for neutral particles, and 1\% for charged particles (the values used in \texttt{CMS-SMP-22-01} \cite{CMS:2024mlf});

    \item 
    turning on/off the effects of hadronization;

    \item 
    turning on/off the effects of multiple-parton interactions (MPI), with hadronization on, as a proxy for the underlying event (UE).
\end{itemize}
We note that, in this simplified context, the EEC is more robust to the effects of particle-level momentum smearing than the mass EWOC in the estimation of \(m_W\), but that the mass EWOC is much more resilient to non-perturbative QCD effects than the EEC.
The presented values for the variations in the EEC are commensurate with corresponding values in the estimation of the top quark mass using the three-point energy correlator, as in \href{https://arxiv.org/pdf/2201.08393\#table.1}{table I} of \Reff{Holguin:2022epo}.
The shifts in the mass EWOC determination of \(m_W\) due to smearing and hadronization are greater in absolute value than those of the determination of \(m_W\) using the mMDT-groomed jet mass.
However, when smearing, hadronization, and MPI are considered together, the overall shift in the peak of the mass EWOC -- and the resulting shift in the determination of \(m_W\) -- is smaller than that of the mMDT-groomed mass distribution.

%=======================
% Outline of the paper
%=======================
The rest of the paper proceeds as follows:
In \Sec{EWOCs}, we define EWOCs and discuss their properties and some applications.
In \Sec{mass}, we examine the phenomenology of the mass EWOC in both proton-proton and electron-positron collisions.
In our study of LHC collisions in \Sec{mass:pp}, we show that the mass EWOC is a particularly powerful tool for extracting mass scales from collision data, using the $W$-boson as an example of a generic hadronically-decaying resonance.
In \Sec{analytic} we present fixed-order results for EWOCs of light-quark jets produced in \(e^+e^-\) collisions and compare them to \pythia{} simulations.
We give concluding thoughts and discuss avenues for future exploration in \Sec{Conclusions}.

% ==============================================
\section{Energy Weighted Observable Correlations (EWOCs)}
% ==============================================
\label{sec:EWOCs}

In this section, we provide an introduction to EWOCs and the physics they probe.
Secs.~\ref{sec:EWOCs:EEC} and \ref{sec:EWOCs:EWOCs} introduce the general considerations and ingredients that go into the construction of EWOCs, presenting EWOCs as flexible, easily interpretable extensions of the more familiar EEC.
Sec.~\ref{sec:EWOCs:IRC} discusses how the use of subjets ensures collinear safety and suppresses non-perturbative effects for generic EWOCs involving non-angular correlations.

% ---------------------------------------
\subsection{A Review of the Energy-Energy Correlator (EEC)}
% ---------------------------------------
\label{sec:EWOCs:EEC}

Qualitatively, the EEC captures information about the angular scales at which energetic pairs of particles tend to strike a detector after a particle collision event.
As visualized in the inset of \Fig{eec:pp_to_ww:with_cartoon}, the EEC at the angular scale \(\chi\) is the average product of the energies of particles separated by an angle \(\chi\), averaged over an ensemble of jets.

More precisely, the EEC is defined as%
\footnote{
    The definition of \Eq{eec_defn} is common in particle phenomenology.
    When all particles are massless, there is an equivalent definition of the EEC in terms of \emph{light-ray operators}: integrals of the stress energy tensor along light rays travelling from the location of the particle collision;
    see, for example, \Reffs{Hofman:2008ar}{Kravchuk:2018htv}.
    The light-ray definition of the EEC may even be applied in the context of conformal field theory, when the concept of a single particle no longer makes sense.
}
\begin{equation}
    \label{eq:eec_defn}
    % \frac{1}{\sigma}\,
    \frac{\dd \Sigma_\text{EEC}}{\dd \chi}
    =
    \frac{1}{\sigma}\,
    \int \dd\sigma \,
    \sum_{
        \text{particles }
        i,\, j
    } \,
    x_i \, x_j \,
    \delta\left(
        \chi \, 
        -
        \theta_{ij}
    \right)
    ,
\end{equation}
where $\chi$ is an angular variable and \(\dd\sigma\) is a differential cross-section with \(\int \dd\sigma = \sigma\).
The subscripts \(i\) and \(j\) denote outgoing particles, \(x_i\) and \(x_j\) describe their energy or transverse momentum fraction and \(\theta_{ij}\) denotes an angular metric.%
\footnote{
    In electron-positron collisions, it is common to replace the \(\delta(\chi-\theta_{ij})\) of \Eq{eec_defn} with \(\delta[\chi - (1-\cos\,\theta_{ij})/2]\).
    \(\theta_{ij}\) denotes the real-space angle between the momenta of particles \(i\) and \(j\).
    The sum on $i,j$ often runs over all particles in an event (in which case it is called an \textit{event shape}), though it can alternatively run over all particles within the set of jets obtained by an appropriate jet definition.
    When treated as an event shape, the EEC energy weights are \(x_i = E_i/E_\text{event}\), while 
    for the jet-based EEC \(x_i = E_i / E_\text{jet}\).
    In proton-proton collisions, the sum runs over all particles in a jet, \(x_i = p_{T,\,i}/p_{T,\,\text{jet}}\), and \(\theta_{ij}\) denotes instead the rapidity-azimuth distance between particles \(i\) and \(j\).
    When computing EECs and general EWOCs in this work, we will always use jet-based definitions.
}

Though the EEC is conceptually simple observable, the use of particles in the definition of the EEC presented in \Eq{eec_defn} makes the EEC somewhat vulnerable to the non-perturbative physics of low-energy QCD, which is notoriously difficult to predict and control, through effects such as hadronization and the underlying event.
The EEC partially mitigates this issue through the use of energy weighting (the factor \(x_i\, x_j\) in \Eq{eec_defn}) such that these low-energy effects are suppressed by their small energy fractions.
The resulting observable is less sensitive than traditional jet observables to the effects of low-energy QCD, but the EEC still uses particles directly, and therefore has not completely eliminated the potential for complicated, non-perturbative effects.

Furthermore, while the angular scales in a particle collision convey a great deal of information about the underlying fundamental physics, they are not able to efficiently characterize everything about the interactions of our microscopic universe.
In order to gather a more rich collection of information about collider physics using energy-weighted correlations, it is tempting to simply change the argument of the delta function in \Eq{eec_defn}, which singles out angular scales.
However, the resulting observable is collinear unsafe (infrared safety is still ensured by the energy weighting).
In the following section, we will introduce EWOCs, which solve this issue by introducing energy-weighted correlations on \textit{subjets}.

% ---------------------------------------
\subsection{Introducing EWOCs}
% ---------------------------------------
\label{sec:EWOCs:EWOCs}

The framework of (subjet-level) energy-weighted observable correlations (EWOCs) that we introduce in this section takes the mitigation of low-energy effects one step further.
In addition to using the energy weighting of the EEC, we use subjets rather than particles;
subjets are constructed by applying a jet algorithm with radius \(\rsub < \Rjet\) to the particles clustered within each jet in an event, and the subjet radius provides an explicit collinear cutoff on contributing radiation.
\footnote{
    Subjet algorithms are not the only regularization schemes which provide a collinear cutoff on the physics of the EEC.
    For example, \Reffs{Barata:2023bhh}{Budhraja:2024xiq} both compute the EEC on collinear-safe collective degrees of freedom based on the Lund plane or on the clustering history of the jet under study, respectively.
}
Any collection of particles which lead to the same subjet configurations will have the same EWOCs, regardless of the details of the particles themselves. 

EWOCs therefore trade the ability to capture correlations between all particles in a jet for the ability to more easily control the effects of low-energy QCD.
As we will see, this tradeoff dramatically increases the richness of the correlations that can be probed by EWOCs.
In addition to the angular correlations probed by the EEC, EWOCs can be used to capture many other correlations -- such as correlations in mass, formation time, or any other observable on subjets -- which are forbidden by collinear safety in the absence of a subjet radius.
An EWOC for a generic observable \(\mathcal{O}(\text{subjet 1, subjet 2})\) on pairs of subjets takes a similar form to \Eq{eec_defn}.
Its definition was given in \Eq{intro_ewoc_def} and is repeated here for convenience:
\begin{equation}
    \label{eq:subjet_ewoc_defn}
    \frac{\dd \Sigma_\mathcal{O}}{\dd \chi}
    =
    \frac{1}{\sigma}
    \int \dd\sigma \,
    \sum_{
        \text{subjets }
        i,\, j
    } \,
    x_i \, x_j \,
    \delta\left(\chi \, - \, \mathcal{O}_{ij}\right)
    \,,
\end{equation}
where the sum over particles has been replaced by a sum over subjets, and the role of \(\theta_{ij}\) has been taken instead by \(\mathcal{O}_{ij} = \mathcal{O}(\text{subjet }i, \text{subjet }j)\), the value of the observable \(\mathcal{O}\) on subjets \(i\) and \(j\).
The normalization $\sigma$ is the jet cross section, and ensures that the EWOC is normalized:
\(\int\! \dd \chi \frac{\dd \Sigma}{\dd\chi} = 1.\)%
\footnote{
    EWOCs can also be defined as event shapes, for which the sum on subjets \(i\) within a particular jet is instead replaced by a sum on jets \(i\) within an entire event.
    However, in this work, we use the jet-based definition of \Eq{subjet_ewoc_defn} for all EWOC computations.
}
The observable \(\mathcal{O}_{ij}\) can depend on the energies and relative angle of subjets \(i\) and \(j\), but also on additional information such as the subjet masses.

The advantage of EWOCs is that they directly probe correlations in an observable of interest.
The choice of subjet definition (e.g.~radius and recombination scheme) determines the set of subjets that contribute to the EWOC, and therefore may also be tuned to capture different physical effects.
For example, the subjet radius may be chosen to capture the subjets emerging from the hadronic decays of a boosted particle, as we explore in \Sec{mass}.
In this work, we focus on the concrete example of the mass EWOC as a proof-of-concept, with a subjet radius chosen to probe decays of the \(W\) boson, but the use of EWOCs is by no means limited to the measurement of mass correlations.
Possible pairwise observables include:
\begin{itemize}
    \item
        The opening angle \(\theta_{ij}\) or the rapidity-azimuth distance \(R_{ij}\) between subjets, which probe angular correlations (used in $e^+e^-$ or $pp$ collisions, respectively).
        These observables lead to the EEC, for which the use of subjets does not provide novel results;%
        \footnote{
            With a subjet radius \rsub{}, the value of the subjet EEC is nearly identical to the particle-level EEC for $\chi>\rsub$, while the collinear cutoff of the subjet radius sets the subjet EEC to zero for $0<\chi<\rsub$.
            The subjet EEC and the particle-level EEC would be exactly the same for \(\chi > \rsub\) if all particles in each subjet were exactly aligned with the subjet direction, with small differences suppressed by the subjet radius size.
        }
        
    \item
        The mass \(m_{ij} = \sqrt{\le(p_i + p_j\ri)^2}\), which probes mass scales%
        \footnote{
            The four-momenta of the subjets depends on the recombination scheme used to define the subjets.
            In this work, we use winner-take-all schemes which enforce that all subjets are massless.
            We have found that this choice of recombination scheme does not affect the performance of the mass EWOC in constraining the $W$-boson mass.
        }, applied to the phenomenolgy of massive decaying particles in \Sec{mass};

    \item
        The formation time \(\tau_{ij} = \max{\le(E_i, E_j\ri)} / m^2_{ij}\), which probes the Lorentz-dilated time scale at which partons (subjets) \(i\) and \(j\)  become ``separate''.%
        \footnote{
            Since formation time is usually associated with partonic splittings, formation time EWOCs may be more useful if the sum on subjet pairs only includes subjets associated with splittings in the branching history of a jet.
        }
        At times  \(t > \tau_{ij}\), partons \(i\) and \(j\) are expected to behave as separated, distinct color charges, while for \(t < \tau_{ij}\) they are expected to behave ``coherently'' as a single color charge.
        We defer the study of formation time EWOCs, especially in the study of heavy-ion collisions, to future work.%
        \footnote{
            The formation time is an especially useful observable in relativistic heavy-ion collisions, where it can probe whether two partons ``split'' before or after the time scale set by the mean-free path of the QGP medium.
            Splittings with formation times much smaller than the mean free path of the medium are therefore expected to be roughly governed by the physics of QCD in empty space/without a medium;
            the formation time at which predictions deviate from QCD without a medium can be used to infer the mean free path of the medium, and therefore how medium effects modify the behavior of QCD splittings~\cite{Gyulassy:1993hr,Baier:1994bd,Zakharov:1996fv,Baier:1996sk,Baier:1996kr,Zakharov:1997uu,Wiedemann:2000ez,Wiedemann:2000za,Gyulassy:2000er,Wang:2001ifa,Kovner:2003zj,Borghini:2005em,Armesto:2007dt,Ovanesyan:2011xy,Ovanesyan:2011kn,Blaizot:2012fh,Fickinger:2013xwa,Apolinario:2014csa,Attems:2022ubu}.
        }
\end{itemize}
Each observable is useful for the study of different physical effects, and each comes with its own behavior under non-perturbative corrections and in analytic computations.

The EEC has traditionally been applied to particle physics in order to extract the strong coupling constant \cite{Martin:1986uq,DELPHI:1990sof,SLD:1994yoe,ATLAS:2015yaa,ATLAS:2017qir,dEnterria:2018cye,Kardos:2018kqj,Ali:2020ksn,dEnterria:2022hzv,Komiske:2022enw,ATLAS:2023tgo,CMS:2024mlf}, and more recently to a broad set of applications including extracting mass scales of boosted objects in colliders \cite{Procura:2022fid,Holguin:2022epo,Holguin:2023bjf,Pathak:2023tmy,Xiao:2024rol,Holguin:2024tkz} and the characterization of relativistic heavy-ion collisions \cite{Lokhtin:2004tx,Lokhtin:2006dp,Andres:2022ovj,Barata:2023zqg,Andres:2023xwr,Yang:2023dwc,Barata:2023vnl,Barata:2023bhh,Barata:2024nqo,Barata:2024ieg}.
More general EWOCs refine and extend the goals of the EEC in particle physics, and the addition of a subjet radius or new pairwise observables may be useful in several contexts.

% -----------------------------------
% IRC Safety Figure:
% -----------------------------------
\begin{figure}[t!]
    \centering
    \subfloat[]{
       \includegraphics[width=0.35\textwidth]{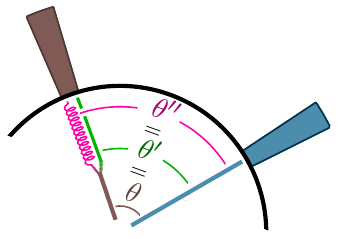}
       \label{fig:EWOCs:cartoon:angle_irc}
    }
    \hspace{2.0cm}
    \subfloat[]{
       \includegraphics[width=0.35\textwidth]{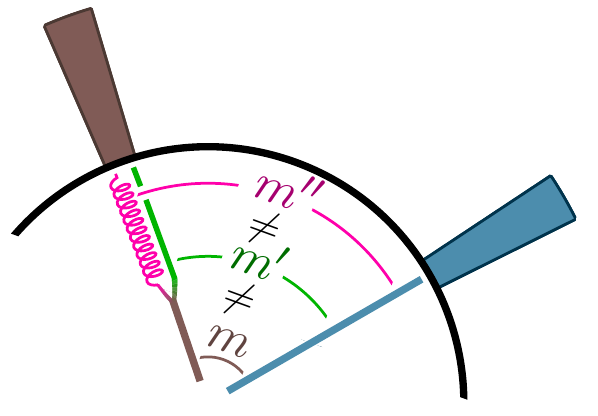}
       \label{fig:EWOCs:cartoon:mass_irc}
    }
    \\
    \subfloat[]{
        \includegraphics[width=.44\textwidth]{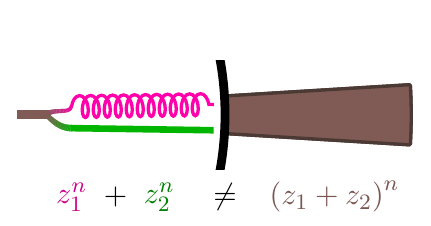}
       \label{fig:EWOCs:cartoon:energyweight_irc}
    }
    % Caption
    \caption{
        Cartoons of collinear splittings (colored lines inside the black hemispheres, for which the brown quark line splits into the pink gluon line and the green quark line) producing the same subjets (colored bars outside of the hemispheres) and the collinear safety properties of
        \hyperref[fig:EWOCs:cartoon:angle_irc]{(a)}
        the EEC (see \Eq{eec_safe}),
        \hyperref[fig:EWOCs:cartoon:mass_irc]{(b)}
        the mass EWOC (see \Eq{m_ewoc_unsafe}), and
        \hyperref[fig:EWOCs:cartoon:energyweight_irc]{(c)} the EWOCs with non-unity energy weights of \Eq{weights_defn}.
        \hyperref[fig:EWOCs:cartoon:angle_irc]{(a)}
        The EEC is collinear safe at the level of both particles and subjets:
        the collinear splitting does not change the set of angles between particles in a jet (see \Eq{eec_safe}).
        \hyperref[fig:EWOCs:cartoon:mass_irc]{(b)}
        The particle-level mass EWOC is collinear unsafe (as are non-angular EWOCs in general) because the set of masses of particle pairs changes after a collinear splitting:
        \(m \neq m' \neq m''\) (see \Eq{m_ewoc_unsafe}).
        The subjet-level mass EWOC, however, is collinear safe as long as subjets are unchanged by collinear splittings.
        For the same reason, \hyperref[fig:EWOCs:cartoon:energyweight_irc]{(c)} generic subjet-level EWOCs remain collinear-safe even in the presence of non-unity energy weights.
    }
    \label{fig:EWOCs:cartoon:particle_irc}
\end{figure}
% -----------------------------------

% ---------------------------------------
\subsection{Infrared/Collinear (IRC) Safety and Non-Perturbative Effects}
% ---------------------------------------
\label{sec:EWOCs:IRC}

As mentioned in \Sec{EWOCs:EWOCs}, particle-level EWOCs -- with \(\rsub = 0\) -- utilizing non-angular observables are collinear unsafe. 
For the special case of the EEC, there is no problem of collinear unsafety:
the particles produced in a collinear splitting still contribute at the exact same angle in the EEC.
However, the values of a \emph{generic} observable that would enter a \emph{particle-level} EWOC differ before and after one of the particles undergoes a collinear splitting.
On the other hand, EWOCs which utilize subjets -- which we will simply refer to as EWOCs -- are manifestly collinear safe as long as the subjet algorithm being used is collinear safe
Furthermore, the use of subjets allows for IRC-safe EWOCs with non-unity energy weights,
\begin{align}
    \label{eq:weights_defn}
    \frac{\dd \Sigma^{(n,m)}_\mathcal{O}}{\dd \chi}
    =
    \frac{1}{\sigma}
    \int \dd\sigma \,
    \sum_{
        \text{subjets }
        i,\, j
    } \,
    x_i^n \, x_j^m \,
    \delta\left(\chi \, - \, \mathcal{O}_{ij}\right)
    \,,
\end{align}
which, for \(n, m > 1\), are even less sensitive to the soft radiation within an event. The benefits of non-unity weights will be discussed below.

The collinear safety of the EEC and the collinear unsafety of the would-be particle-level mass EWOC is illustrated in \Fig{EWOCs:cartoon:particle_irc}, where subjets are visualized as bars on the outside of the circles which are unchanged under a collinear splitting.
Fig.~\ref{fig:EWOCs:cartoon:particle_irc} focuses on the simple example of a two-particle final state and the corresponding three-particle final state after a collinear splitting.
Fig.~\ref{fig:EWOCs:cartoon:angle_irc} visualizes the collinear safety of the particle-level EEC, contrasted against the collinear unsafety of generic, particle-level EWOCs, such as the mass EWOC visualized in \Fig{EWOCs:cartoon:mass_irc}.
Fig.~\ref{fig:EWOCs:cartoon:energyweight_irc} visualizes the collinear unsafety of generic energy weighting factors in would-be particle-level EWOCs, including the particle-level EEC.
While the use of non-unity energy weights is collinear unsafe at particle-level, the subjets used in the computation of EWOC are unchanged under collinear splittings, as long as the subjet algorithm is collinear safe.

To understand the collinear safety of the EEC, and the unsafety of the particle-level mass EWOC, more precisely, let us denote the momentum fraction of the right (blue) particle by \(z_2\), the left (brown) particle before a collinear splitting by \(z\), and the particles emerging from the collinear splitting (green and pink) by \(z'\) and \(z''\).
The pairwise angles between the right particle and either the collinear parent and collinear-split children are \(\theta,\theta'\) and \(\theta''\), respectively, and the associated pairwise masses are \(m, m',\) and \(m''\).
Since the splitting is collinear, we have \(\theta = \theta' = \theta''\) as well as \(z = z' + z''\)
Already, we find that the associated pairwise correlations captured by the EEC do not change under a collinear splitting\footnote{
    Similar arguments hold for any number of particles, as well as for \textit{contact terms} -- the singular contributions at \(\chi = 0\) due to particle self-correlations:
    \(
        \dd \Sigma_\text{\tiny EEC}/\dd \chi
        \supset
        z^2
        \, \delta\le(\chi\ri)
        =
        \le(z^{\prime\,2} + z^{\prime\prime\,2} + 2 z^\prime z^{\prime\prime} \ri)
        \, \delta\le(\chi\ri)
        \subset
        \dd \Sigma_\text{\tiny EEC}/\dd \chi\,\big|_\text{split}
        \,.
    \)
}:
\begin{align}
    \label{eq:eec_safe}
    \frac{\dd \Sigma_\text{\tiny EEC}}{\dd \chi}
    \supset
    2 \, z_2 \, z
    \, \delta(\chi - \theta)
    \,\,\,
    =
    2 \, z_2 \, \le(
        z'
        \,
        \delta(\chi - \theta')
        +
        z''
        \,
        \delta(\chi - \theta'')
    \ri)
    \subset
    \frac{\dd \Sigma_\text{\tiny EEC}}{\dd \chi}\biggr|_\text{split}
    \,.
\end{align}
For the would-be particle-level mass EWOC, however, the collinear splitting changes the value of the distribution
\begin{align}
    \label{eq:m_ewoc_unsafe}
    \frac{\dd \Sigma_{m}}{\dd \chi}
    \supset
    2 \, z_2 \, z
    \, \delta(\chi - m)
    \,\,\,
    \boldsymbol{\neq}
    2 \, z_2 \, \le(
        z'
        \,
        \delta(\chi - m')
        +
        z''
        \,
        \delta(\chi - m'')
    \ri)
    \subset
    \frac{\dd \Sigma_{m}}{\dd \chi}
    \biggr|_\text{split}
    \,.
\end{align}
Neither \(m'\) nor \(m''\) is equal to \(m\), and the support of the distribution has changed under a collinear splitting.
We conclude that the would-be particle-level EWOC is collinear-unsafe.

The use of energy weights larger than one, as defined in \Eq{weights_defn}, has the potential to further reduce the effects of additive contamination such as pileup and the overwhelming underlying event in relativistic heavy-ion collisions \cite{Barata:2023bhh}.%
\footnote{
    Ref.~\cite{Barata:2023bhh} provides an earlier discussion of how a different collinear regularization of the EEC facilitates the IRC-safe use of non-unity energy weights, and notes that the use of higher energy weights may provide a way to mitigate the low-energy effects of the formidable underlying event of relativistic heavy-ion collisions.
}
In \Sec{mass:pp}, we restrict to the case \(n = m\) and examine how energy weights can be used to mitigate non-perturbative effects and yield more robust results for EWOCs.%
\footnote{
    We do not consider the case \(n \neq m\), though it may be interesting to do so e.g.~to pick out configurations with one energetic and one softer subjet, or when using asymmetric pairwise observables \(\mathcal{O}_{ij} \neq \mathcal{O}_{ji}\).
    The latter in particular may be interesting in the study of EWOCs characterizing formation times.
}
We note, however, that the additional flexibility and infrared stability granted by the energy weights come at the expense of losing a simple sum rule:
the integral over $\chi$ of \(
    \dd \Sigma^{(n,m)}_\mathcal{O} / \dd \chi
\) will no longer yield one unless \(n = m = 1\).%
\footnote{
    It is also possible to define EWOCs with higher energy weights which nonetheless integrate to one, by ensuring that the sum of each energy weighting factor over subjets gives one;
    for example, one could replace the energy-weighting factor \(x_i^n\) in \Eq{weights_defn} with \(x_i^n / \sum_k x_k^n\), and similarly replace \(x_j^m\) by \(x_j^m / \sum_k x_k^m\).
    However, these EWOCs are theoretically more complicated because the denominator $\sum_k x_k^n$ depends sensitively on \emph{all} subjets. 
    This definition also suffers from greater hadronization and underlying event corrections, and thus does not offer improvements in the main application discussed here -- the estimation of \(m_W\) using the mass EWOC.
    We thank Jesse Thaler for discussions on this point.
}

Fortunately, the use of subjet radii leads to results which are not only collinear-safe, but also more stable to non-perturbative effects.
On physical grounds, this is because subjets are collections of particles that approximate high-energy partons produced during a hard process, while hadronization is a low-energy phenomenon.
More technically, a subjet radius is a type of infrared cutoff:
in a scattering process of characteristic energy scale \(Q\), subjets with \(\rsub\ll 1\) contained within jets of radius \Rjet{} probe physics at scales between \(Q\,\Rjet\) and \(Q\rsub\).
Therefore, if \(Q\rsub\gg\Lambda\), where \(\Lambda\) is an energy scale of non-perturbative physics, we are in a regime where perturbation theory is valid and non-perturbative effects are suppressed by powers of \(\Lambda/(Q\rsub)\).

For example, hadronization effects in the mass EWOC are controlled by powers of \(\lqcd/(Q \rsub)\) in the regime of a small subjet radius.
In the regime \(Q \sim Q \Rjet \gg m \gg Q \rsub \gg \lqcd\), where \(m\) is the argument of the mass EWOC, there are also hadronization effects which scale with \(\lqcd/Q\) and \(\lqcd/m\) and are suppressed relative to the leading hadronization corrections, which scale with \(\lqcd/(Q \rsub)\).
Similar conclusions hold for a generic EWOC when \(m\) is replaced by the relevant scale.

% ==============================================
\section{Mass Extraction with EWOCs}
% ==============================================
\label{sec:mass}

In this section, we examine EWOCs in greater depth by focusing on the mass EWOC.
In \Sec{mass:pp}, we apply EWOCs to one of the most important contexts for jet substructure:
the characterization of heavy, boosted objects that emerge in proton-proton collisions.
In particular, we demonstrate how the EWOC framework can be used to extract the mass of hadronically-decaying boosted \(W\) bosons created in proton-proton collisions, with results that are robust to non-perturbative corrections.
In \Sec{analytic} we explore the fixed order and leading-logarithmic structure of mass EWOCs in the theoretically simpler context of light-quark jets produced in electron-positron collisions, and compare them to results obtained with \pythia{}.

% -------------------------------------
\subsection{Observables and Samples}
% ---------------------------------------
The data presented in this work are derived from sample events generated using \pythia{} \cite{Bierlich:2022pfr} with the default Monash tune \cite{Skands:2014pea}, with jets and subjets clustered using \texttt{FastJet} \cite{Cacciari:2011ma}.
The code used in our analysis is available at \texttt{ResolvedEnergyCorrelators}~\cite{Alipour-fard:2024szj,github:RENC}. 

In our studies of proton-proton collisions at \(\sqrt{s}=\,\)14 TeV in \Sec{mass:pp}, we cluster jets using the anti-\(k_t\) algorithm \cite{Cacciari:2008gp} with \(\Rjet = 0.8\) and cluster subjets using the \(k_t\) algorithm \cite{Catani:1993hr,Ellis:1993tq}, and use the winner-take-all (WTA) \(p_T\) recombination scheme \cite{Bertolini:2013iqa,Larkoski:2014uqa} for both jets and subjets.
We restrict our analysis to jets with a maximum pseudorapidity of \(|\eta| < 4\) and a minimum transverse momentum of \(p_T > 500\,\)GeV unless otherwise stated, ensuring that the associated \(W\) bosons are sufficiently boosted for their decay products to lie within a single jet.
The anti-\(k_t\) jet algorithm ensures that our jet boundaries are insensitive to the presence of contaminating radiation such as the underlying event.
The \(k_t\) subjet algorithm ensures that our subjets are more likely to be correlated with the hard sub-prongs (quarks) of our \(W\) jets.%
\footnote{
    The Cambridge-Aachen (C/A) \cite{Dokshitzer:1997in,Wobisch:1998wt} algorithm, for example, is more likely to cluster the hard, narrow core and the soft, wide-angle radiation produced by a single quark into separate subjets \cite{Atkin:2015msa,istep:2016,boost:2024}.
    However, contributions of these spurious, low-energy C/A subjets are suppressed by the energy-weighting of the EWOC, and we find very similar results for EWOCs using both \(k_T\) and C/A subjets.
}
Using a WTA recombination scheme ensures that the mass EWOC only captures correlations between distinct subjets, and is insensitive to the masses of individual subjets;%
\footnote{
    Recombination schemes that produce massive subjets may also offer interesting applications.
    For example, in the \(E\)-scheme for subjet recombination \cite{Blazey:2000qt}, the mass EWOC interpolates between collinear-unsafe mass correlations -- the particle-level mass EWOC with \(\rsub=0\) -- and the jet mass itself -- when \(\rsub=\Rjet\).
}
our results for groomed mass distributions instead uses the standard \(E\)-scheme for jet recombination \cite{Blazey:2000qt}.

In our studies involving \(e^+\,e^-\) collisions at \(\sqrt{s}=\,\)1 TeV in \Sec{analytic}, we cluster jets and subjets with the angular-ordered, electron-positron version of the Cambridge-Aachen (C/A) algorithm
\cite{Dokshitzer:1997in,Wobisch:1998wt}.
The use of the C/A algorithm is more appropriate for the comparison of \pythia{} with our analytic results, since the branching structure of the angular-ordered tree of emissions produced by the C/A algorithm mimics the angular-ordered structure of perturbative QCD \cite{Ellis:1996mzs}.
We study jets with \(\Rjet = 1\) and use the WTA \(|p|\) recombination scheme \cite{Larkoski:2014uqa} for both jets and subjets, again ensuring that all subjets are massless.
We also restrict our analysis to jets with a minimum energy of \(E > 100\,\)GeV in order to avoid contributions from extremely soft jets and thereby facilitate comparisons to analytic results.%
\footnote{
    For EWOC computations, we define energy fractions as the ratio of the subjet energy to the energy of the full jet.
    Therefore, while the energy weighting suppresses contributions from parametrically soft subjets, it does not suppress contributions from soft \textit{jets}.
    The energy cut ensures that our EWOCs are not entirely swamped by soft jets.
    Since \(e^+ e^-\) collisions usually contain two hard jets of energy \(E \sim \sqrt{s}/2 \gg 100\,\)GeV, our results are fairly insensitive to changes in this choice of energy cut.
}

% --------------------------------------
\subsection{Mass at the LHC: \texorpdfstring{\(p p\to W^+ W^-\)}{p p to W+ W-}}
% ---------------------------------------
\label{sec:mass:pp}

% -----------------------------------
% LHC "Money" Plot:
% -----------------------------------
\begin{figure}[t!]
    \centering
    \subfloat[]{
        \centering
        \includegraphics[width=.48\textwidth]{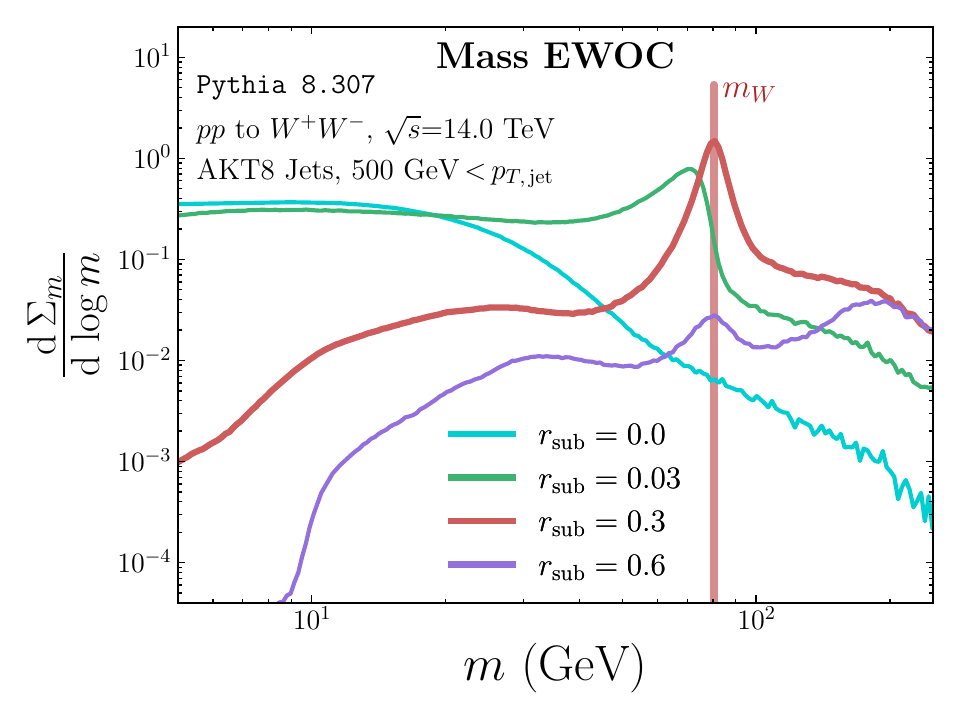}
        \label{fig:money:rsubs}
    }
    \subfloat[]{
        \centering
        \includegraphics[width=.48\textwidth]{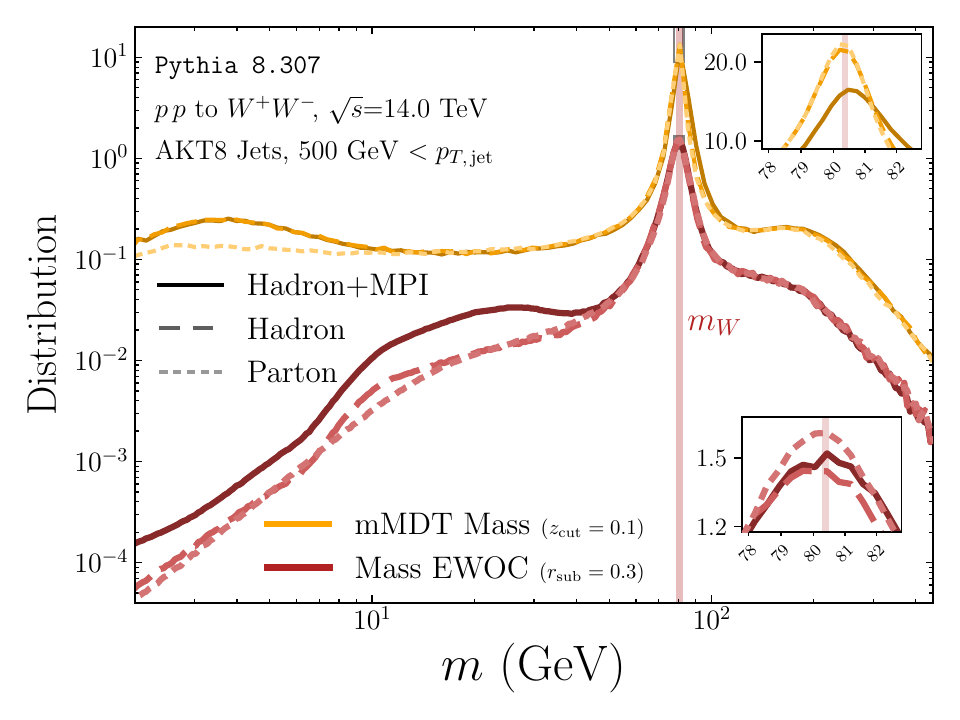}
        \label{fig:money:mpi}
    }
    % Caption
    \caption{
        Mass EWOCs for \(W\)-boson pair production at the LHC at \(\sqrt{s} = 14\) TeV for anti-\(k_T\) jets with \(k_T\) subjets.
        \hyperref[fig:money:rsubs]{(a)} 
        Mass EWOCs for several subjet radii \rsub{};
        the peak at the $W$ mass is most pronounced if \rsub{} is near the mean angular separation between the decay products of the \(W\) boson, \(\Delta \theta \sim 0.3\).
        \hyperref[fig:money:mpi]{(b)} 
        The mass EWOC for \(\rsub = 0.3\) compared to the distribution of the Soft-Drop-groomed \(W\)-jet mass.
        The mass EWOC near \(m_W\) is more robust to the presence of the underlying event (multiple parton interactions) than the groomed jet mass, though it experiences large corrections due to UE in the small-mass region.
    }
    \label{fig:pp_to_ww:money}
\end{figure}
% -----------------------------------

% -----------------------------------
% Comparing Subjet Radii: p p to W W
% -----------------------------------
\begin{figure}[t!]
    \centering
    \subfloat[]{
        \includegraphics[width=.48\textwidth]{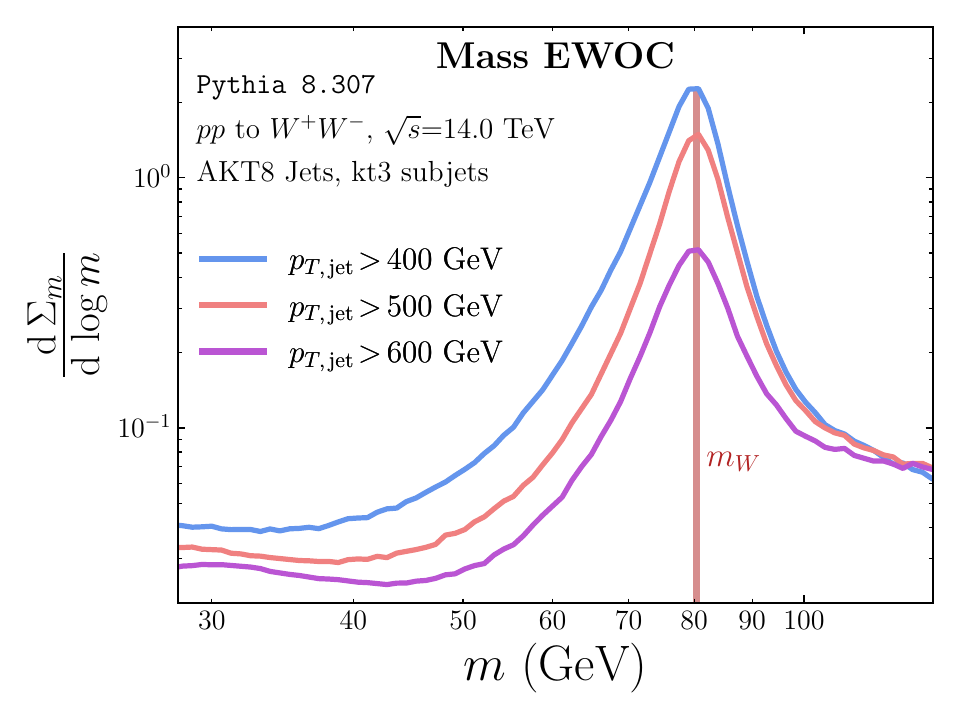}
        \label{fig:m_ewoc:pp_to_ww:compare-pT}
    }
    % Caption
    \subfloat[]{
        \includegraphics[width=.48\textwidth]{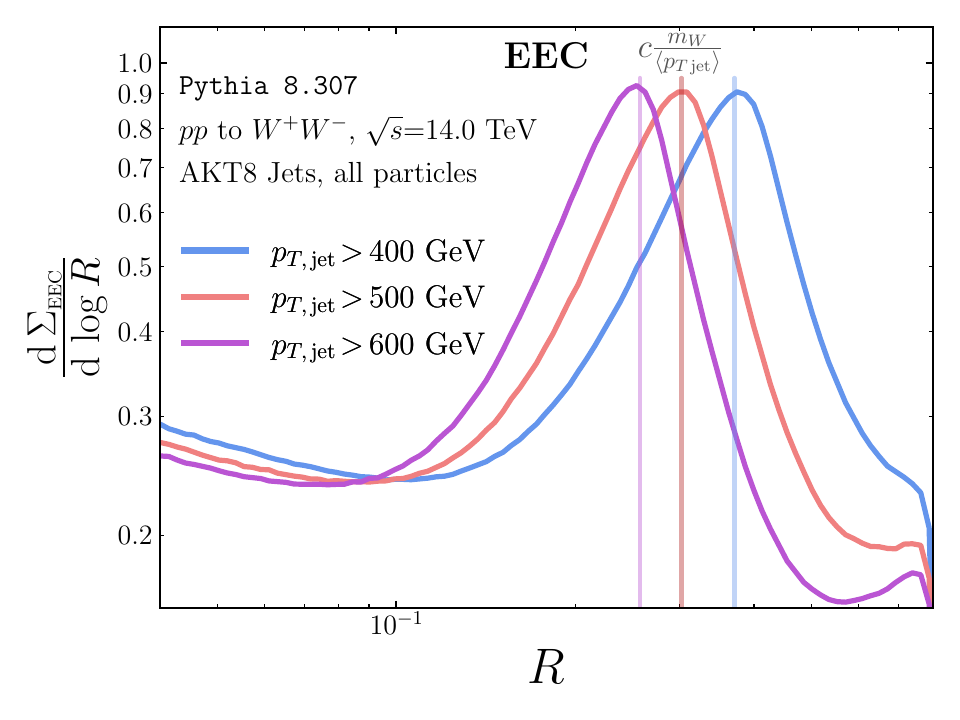}
        \label{fig:eec:pp_to_ww:compare-pT}
    }
    \caption{
        Variation in
        \hyperref[fig:m_ewoc:pp_to_ww:compare-pT]{(a)} 
        the mass EWOC with \(\rsub=0.3\) and
        \hyperref[fig:eec:pp_to_ww:compare-pT]{(b)} 
        the EEC, for simulated LHC \(W\)-boson pair production events as the cut on the minimum jet \(p_T\) is varied.
        The peak position of the mass EWOC is invariant to the choice of minimum \(p_T\) of the \(W\) jets, while the peak position of the EEC changes as the minimum \(p_T\) is varied.
        Nonetheless, the peak of the EEC is well approximated by \(R_\text{peak} \approx \, c \,\,  m_W / \le\langle p_{T,\,\text{jet}}\ri\rangle\), where \(\le\langle p_{T,\,\text{jet}}\ri\rangle\) is a function of the minimum \(p_T\) cut and \(c \sim 2.5\) is the same for each value of the minimum \(p_T\).
    }
    \label{fig:pp_to_ww:compare-pts}
\end{figure}
% -----------------------------------

{
% -----------------------------------
\begin{figure}[t!]
    % \vspace{-1.5cm}
    \centering   
    \subfloat[]{
        \includegraphics[width=.48\textwidth]{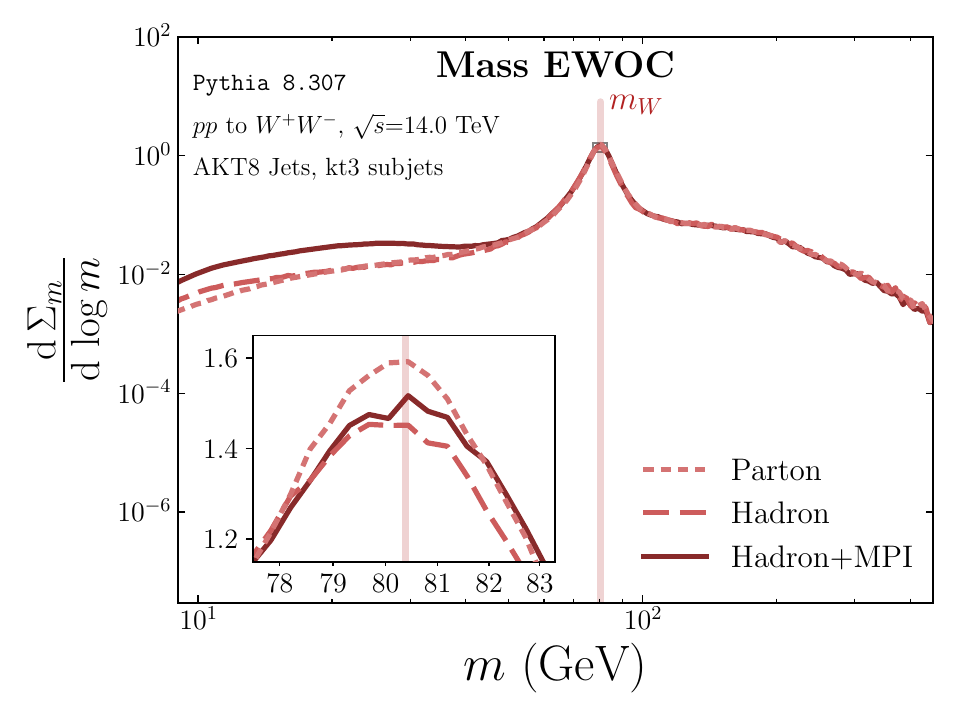}
        \label{fig:m_ewoc:p_v_h_v_mpi:rsub_3}
    }
    \subfloat[]{
        \includegraphics[width=.48\textwidth]{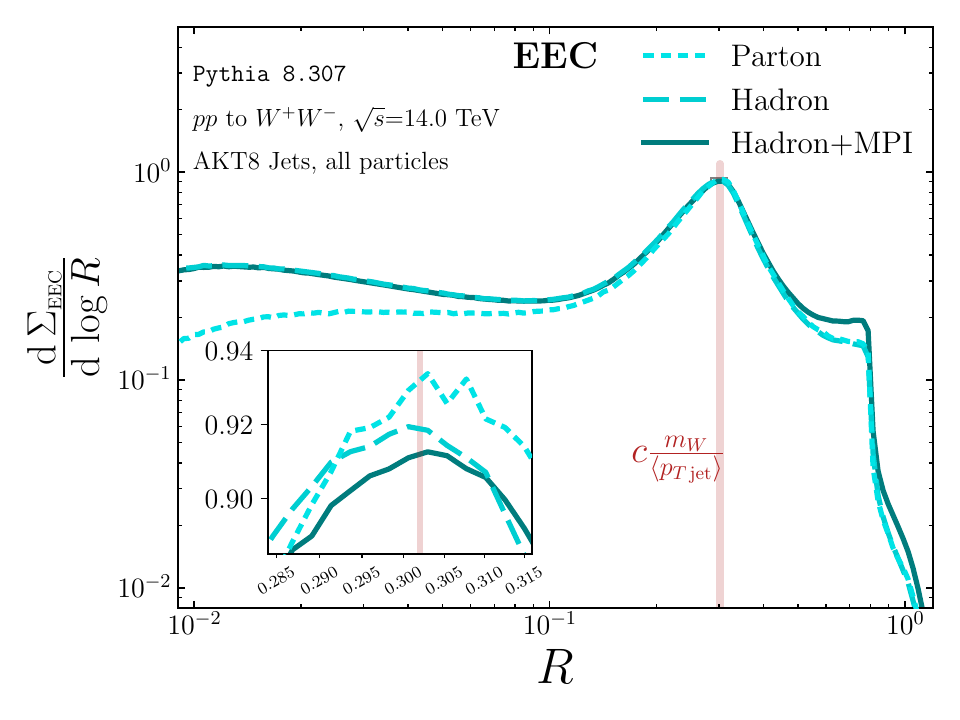}
        \label{fig:eec:p_v_h_v_mpi}
    } 
    \caption{
        Non-perturbative hadronization and UE (MPI) effects on
        \hyperref[fig:m_ewoc:p_v_h_v_mpi:rsub_3]{(a)} 
        the mass EWOC with \(\rsub=0.3\), and
        \hyperref[fig:eec:p_v_h_v_mpi]{(b)} 
        the EEC.
        Both have peaks which are resilient to each source of non-perturbative corrections.
    }
    \label{fig:pp_to_ww:14TeV:p_v_h_v_mpi}
\end{figure}
% -----------------------------------
}

% -----------------------------------
\begin{figure}[t!]
    \centering
    \subfloat[]{
        \includegraphics[width=.48\textwidth]{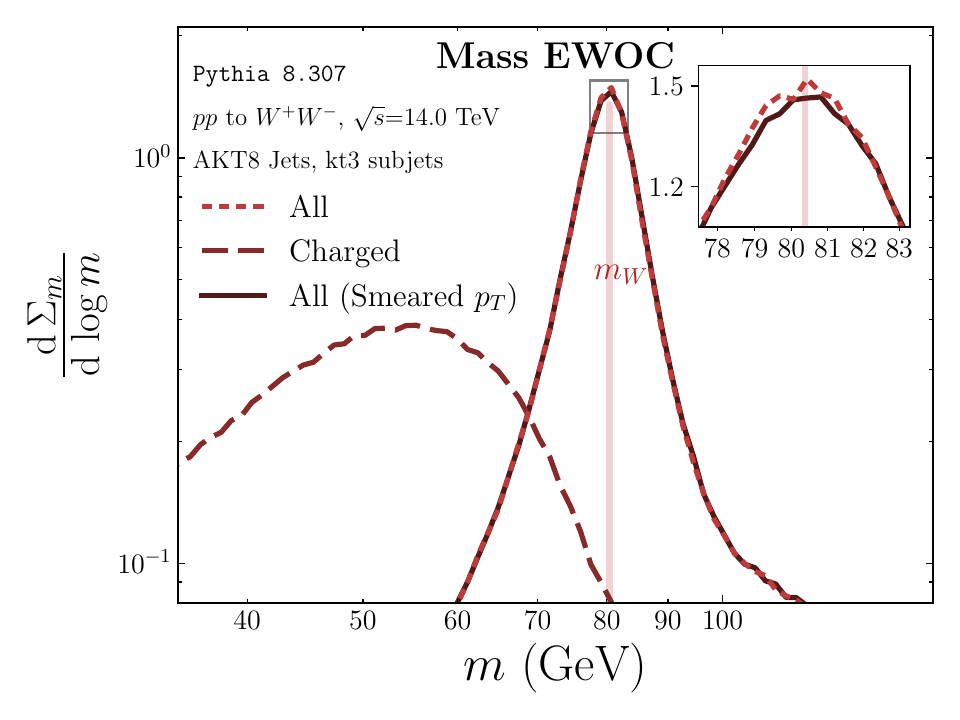}
        \label{fig:m_ewoc:rsub_3:smearing}
    }
    \subfloat[]{
        \includegraphics[width=.48\textwidth]{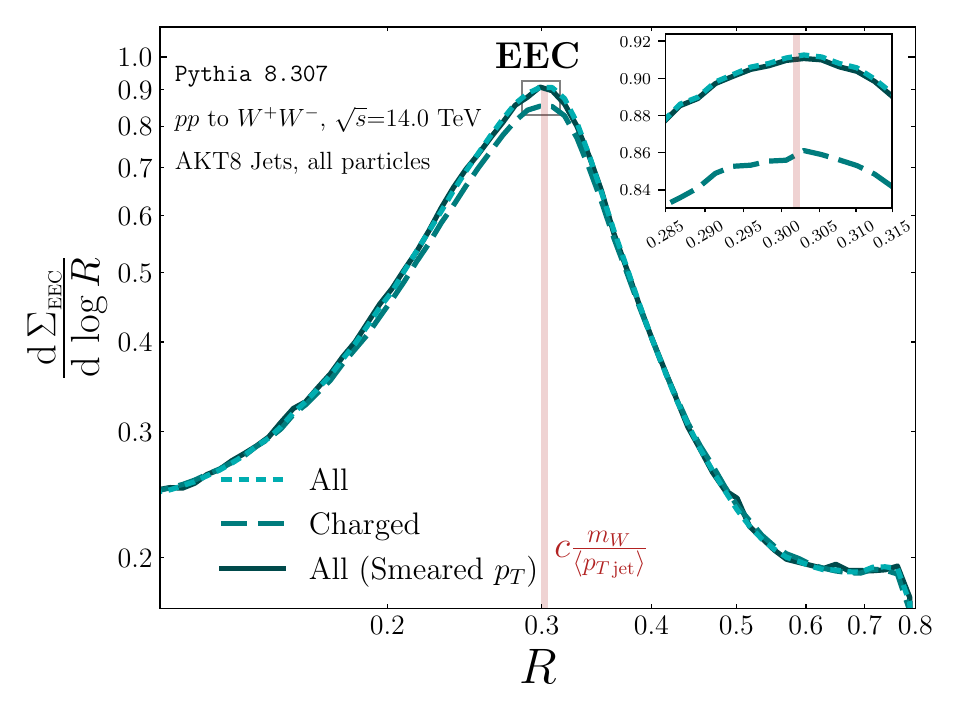}
        \label{fig:eec:smearing}
    }
    \caption{
        Effects of the exclusion of neutral particles, and of a Gaussian model of particle-level momentum smearing emulating the analysis of \texttt{CMS-SMP-22-015} \cite{CMS:2024mlf}, on
        \hyperref[fig:m_ewoc:rsub_3:smearing]{(a)} 
        the mass EWOC with \(\rsub=0.3\), and
        \hyperref[fig:eec:smearing]{(b)} 
        the EEC.
        While both the EEC and the mass EWOC are robust to the effects of momentum smearing, only the EEC is qualitatively unchanged by the exclusion of neutral particles.
        The mass EWOC, on the other hand, is roughly rescaled by a factor of 2/3 when neutral particles are ignored.
    }
    \label{fig:pp_to_ww:14TeV:smearing}
\end{figure}
% -----------------------------------

% -----------------------------------
\begin{figure}[t!]
    \centering
    \subfloat[]{
        \includegraphics[width=.48\textwidth]{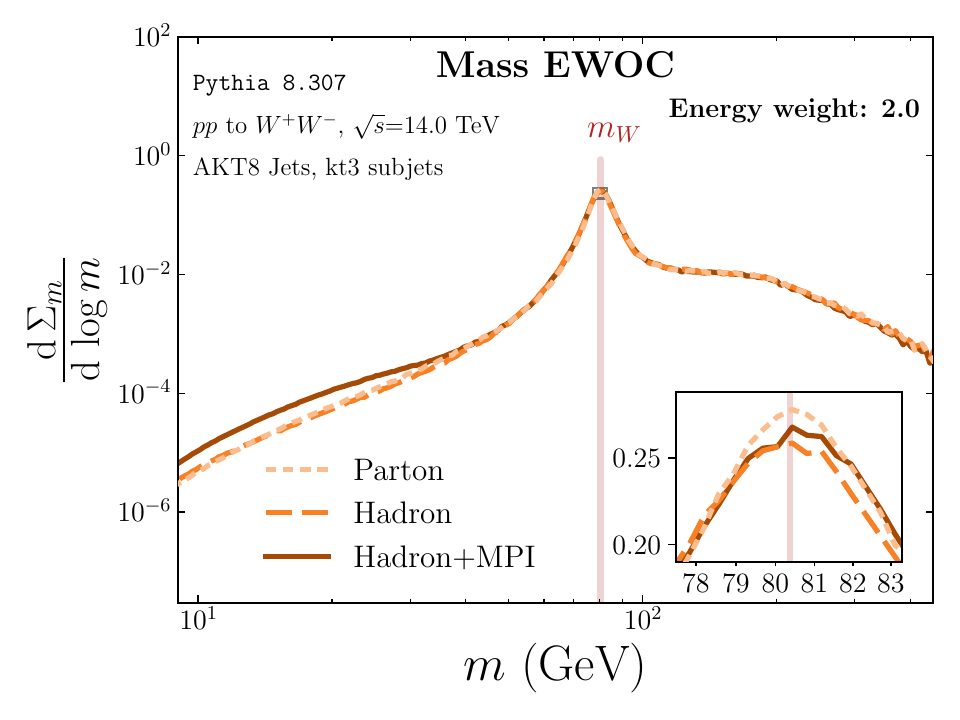}
        \label{fig:m_ewoc:rsub_3:weight_2}
    }
    \subfloat[]{
        \includegraphics[width=.48\textwidth]{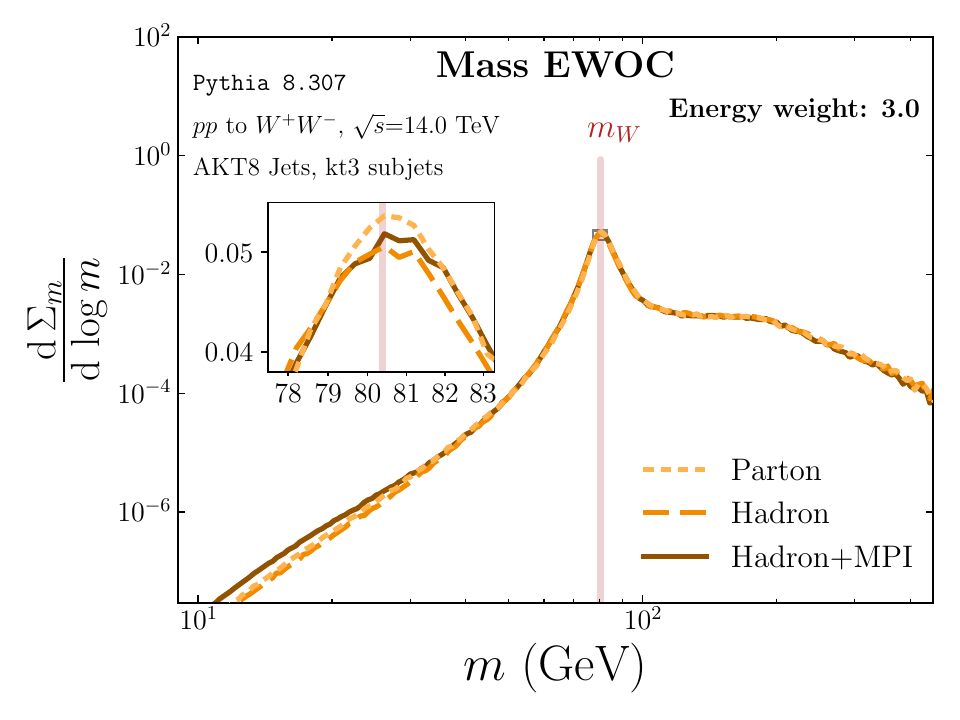}
        \label{fig:m_ewoc:rsub_3:weight_3}
    }
    \caption{
        Non-perturbative hadronization and underlying event corrections to the mass EWOC with \(\rsub=0.3\) and non-unity energy weights \(n > 1\) to mitigate low-energy effects (see \Sec{EWOCs:IRC}), for
        \hyperref[fig:m_ewoc:rsub_3:weight_2]{(a)}
        \(n=m=2\) and
        \hyperref[fig:m_ewoc:rsub_3:weight_3]{(b)}
        \(n=m=3\).
        Notably, larger energy weights mitigate the effects of UE at low mass scales.
        Both distributions, like the mass EWOC with energy weight of unity, are resilient to hadronization effects.
    }
    \label{fig:pp_to_ww:14TeV:weights}
\end{figure}
% -----------------------------------

\begin{table}
    \vspace{-1.0cm}

    % Table
    \begin{tabular}{|p{2.3cm}||p{1.9cm}|p{1.9cm}|p{1.9cm}|}
     \hline
     \multicolumn{4}{|c|}{
     \centering
       \textbf{Shift in \(\boldsymbol{m_W}\) Determination}
     }
     \\
     \multicolumn{4}{|c|}{
       \centering
       from the peak of each distribution
     }
     \\
     \hline
     \centering
     \vspace{0pt}
     \(\boldsymbol{\Delta}\)
     \vspace{0pt}
     &
     \ccw{
     \parbox[m]{\columnwidth}{
        \vspace{12pt}
        \hspace{5pt}
        \(\boldsymbol{n=1}\)
    }
     }
     &
     \ccww{
     \parbox[m]{\columnwidth}{
         \vspace{12pt}
        \hspace{5pt}
         \(\boldsymbol{n=2}\)
     }
     }
     & 
     \ccwww{
     \centering
     \parbox[m]{\columnwidth}{
        \vspace{12pt}
        \hspace{5pt}
        \(\boldsymbol{n=3}\)
    }
     }
     \\
     \hline
     \hline
     \centering
     \vspace{-0.2pt}
     Smearing

     {\small{(cf \Reff{CMS:2024mlf})}}
     \vspace{10.5pt}
     &
     \vspace{8pt}
     \hspace{5pt}
     \ccw{\wsmear{} MeV}
     &
     \vspace{8pt}
     \hspace{5pt}
     \ccww{\wwsmear{} MeV}
     &
     \vspace{8pt}
     \hspace{5pt}
     \ccwww{\wwwsmear{} MeV}
     \\
     \hline
     \centering
     \vspace{0pt}
     Parton vs.

     Hadron
     \vspace{10pt}
     &
     \vspace{8pt}
     \hspace{5pt}
     \ccw{\whad{} MeV}
     &
     \vspace{8pt}
     \hspace{5pt}
     \ccww{\wwhad{} MeV}
     &
     \vspace{8pt}
     \hspace{5pt}
     \ccwww{\wwwhad{} MeV}
     \\
     \hline
     \centering
     \vspace{0pt}
     UE (MPI)
     
     On/Off
     \vspace{10pt}
     &
     \vspace{8pt}
     \hspace{5pt}
     \ccw{\wue{} MeV}
     &
     \vspace{8pt}
     \hspace{5pt}
     \ccww{\wwue{} MeV}
     &
     \vspace{8pt}
     \hspace{5pt}
     \ccwww{\wwwue{} MeV}
     \\
     \hline
    \end{tabular}
    \hspace{-18pt}
    \raisebox{-4.880cm}{
    \begin{tikzpicture}
    \begin{axis}[
        xbar,
        axis lines=left, 
        xlabel={\(\abs{\boldsymbol{\Delta}}\quad\)(GeV)},
        height=7.170cm,
        width=0.48\textwidth,
        xmin=0, xmax=.3,
        ymin=0,
        ytick={0,24,48,72},
        extra y ticks={0,24,48,72},
        ymax=75,
        yticklabels={},
        extra y tick labels={}, 
        extra y tick style={grid=major},
        legend style={
         at={(0.48,1.35)},
         anchor=north,
         legend columns=1,
         /tikz/every even column/.append style={row sep=0.2cm}},
        y axis line style=-,
        ]
    \addplot [color=wcol!90!black, fill=wcol!15]
    coordinates {
        (abs{\wsmear}/1000,72) 
        (abs{\whad}/1000,48)
        (abs{\wue}/1000,24)
    };
    \addplot [color=wwcol!60!wcol!90!black, fill=wwcol!30]
    coordinates {
        (abs{\wwsmear}/1000,60) 
        (abs{\wwhad}/1000,36)
        (abs{\wwue}/1000,12)
    };
    \addplot [color=wwwcol!30!wwcol!90!black, fill=wwwcol!35]
    coordinates {
        (abs{\wwwsmear}/1000,48) 
        (abs{\wwwhad}/1000, 24)
        (abs{\wwwue}/1000,0)
    };
    
    \legend{\(n=1\), \(n=2\), \(n=3\)};

    \end{axis}
    
    % Adding arrow and text
    \node[anchor=west] (source) 
        at (4.9cm,5.5cm){};
    \node (destination)
        at (0.45cm,5.5cm){};
    \draw[{Round Cap}-{Latex[round]}, very thick, red!80!black](source)--(destination); 
    \node[
      above=-0.0cm of destination, 
      xshift=.5cm,
      text width=1.5cm,
      align=left
    ] {\textcolor{red!60!black}{More robust}};
    \node[
      above=0.0cm of source, 
      xshift=-.4cm,
      text width=1.5cm,
      align=right
    ] {\textcolor{red!65!blue}{Less robust}};
    \end{tikzpicture}
    }  
    \caption{
        Shifts in the peaks of mass EWOCs with different energy weightings \(n\) due to detector effects or non-perturbative physics, presented in the style of table \ref{tab:obs_comparison}.
        We observe that the determination of \(m_W\) is less robust to particle-level momentum smearing when using higher energy weights, but that higher energy weights are more robust to non-perturbative QCD effects.
    }
    \label{tab:energyweights}
\end{table}

We now use mass EWOCs to extract the mass of the \(W\) boson from \(pp \to W^+ W^-\) events generated in \pythia{} at parton-level, hadron-level, and in the presence of the underlying event (UE) produced by multiple parton interactions (MPI).
The two-pronged hadronic decay of the \(W\) boson, \(W \to q \,\overline{q}\), is particularly well suited for study with the pairwise (or two-point) mass EWOC:
subjets are proxies for the quarks emerging from the \(W\) decay, and the pairwise subjet mass becomes a proxy for the mass of the \(W\) boson itself.
Furthermore, while we focus on the \(W\) boson in this section, similar methods may be used for the analysis of generic two-pronged hadronic decays, such as hadronic two-prong decays of a hypothetical beyond-the-standard-model $Z'$.

Fig.~\ref{fig:pp_to_ww:money} shows the mass EWOC at several different values of the \(k_T\) subjet radius as well as a direct comparison of the mass EWOC to the mass distribution of jets groomed with the modified mass drop tagger (mMDT) \cite{Dasgupta:2013ihk,Larkoski:2014wba}, which is much more robust than the ungroomed jet mass in the estimation of \(m_W\).
The mass EWOC is best at singling out the \(W\)-boson mass when we pick a subjet radius roughly equal to the expected separation between the partonic decay products of the \(W\):
\(r_\text{sub} \sim m_W / \le\langle p_{T\,\text{jet}}\ri\rangle\sim 0.3\).
At this tuned subjet radius, the location of the mass EWOC peak, and the corresponding inference of the \(W\) mass, is extremely robust to hadronization and UE.
We also note that the mass EWOC at low masses (away from the peak at \(m_W\)) is more robust to hadronization than the mMDT-groomed jet mass distribution.
However, since subjets capture all additive contamination due to the UE within the subjet radius, the mass EWOC gains larger UE corrections at low masses than the mMDT-groomed jet mass.

Fig.~\ref{fig:pp_to_ww:compare-pts} visualizes the effects of changing the selection cuts on the \(W\)-jet samples by varying the minimum \(p_T\) of the \(W\)-jets by 100 GeV about \(p_{T,\,\text{min}} = 500\) GeV.
Though the mass EWOC has a peak that remains at the \(W\)-mass for each value of \(p_{T,\,\text{min}}\), the peak of the EEC shifts as one varies the minimum \(p_T\):
changing the allowed \(p_T\) of the jets also changes the associated angular scales between their constituents.

In \Figs{pp_to_ww:14TeV:p_v_h_v_mpi}{pp_to_ww:14TeV:weights}, we examine non-perturbative corrections to EWOCs through the effects of hadronization and UE.
Fig.~\ref{fig:pp_to_ww:14TeV:p_v_h_v_mpi} compares non-perturbative effects in the mass EWOC and the EEC, focusing on the changes in the peak of each distribution.
The peak of both the mass EWOC and the EEC remain nearly unchanged, and have the potential to provide robust determinations of \(m_W\).
Away from the peak, however, both distributions are affected by non-perturbative physics.
At small angular scales, the EEC receives relatively large corrections from hadronization but relatively small corrections from UE:
hadronization has the potential to change angular scales of hard particles within a jet, while UE provides a background of soft particles which are damped by the energy weighting of the EEC.
On the other hand, at small mass scales, the mass EWOC is unchanged by hadronization but receives relatively large corrections from UE:
subjets comprise collective degrees of freedom which are relatively unchanged by hadronization by construction, but which gain contributions from UE to their momenta -- and therefore their pairwise masses -- proportional to the subjet area.

In \Fig{pp_to_ww:14TeV:smearing}, we show changes in the mass EWOC and the EEC due to the exclusion of neutral particles as well as a rough model of experimental detector effects.%
\footnote{
    The EEC can naturally be extended to measurements on charged particles using the track function formalism \cite{Chang:2013rca,Chang:2013iba,Chen:2020vvp,Li:2021zcf,SchrijndervanVelzen:2022ftm,Jaarsma:2022kdd,Lee:2023xzv,Jaarsma:2023ell,Barata:2024nqo}, and similar developments for EWOCs may benefit applications relying on the superior resolution of track-based measurements.
}
To model detector effects, we implement a particle-level Gaussian momentum smearing of \(5\%\) for neutral particles, \(3\%\) for photons, and \(1\%\) for charged particles, using the values of the analysis in \texttt{CMS-SMP-22-015} \cite{CMS:2024mlf}.
We do not implement any angular smearing, however, citing the excellent angular resolution of the CMS detector \cite{CMS:2024mlf,Holguin:2024tkz}.
We find that the peak of the mass EWOC and of the EEC receive negligible contributions from the smearing of particle momenta.
Furthermore, the EEC remains roughly unchanged by the exclusion of neutral particles.
However, the exclusion of neutral particles changes the masses of subjets and hence the shape of the mass EWOC, rescaling each by roughly a factor of \(2/3\).\footnote{
    This rescaling of energy scales when excluding neutral particles can be argued physically in the ``isospin limit'':
    in the case where all outgoing particles are pions, and the \(\pi^+\), \(\pi^-\), and \(\pi^0\) are all present with similar properties in the final state of a collision, ignoring the neutral pion leads to a loss of \(\sim 1/3\) of the energy of each event.
}.

Finally, in \Fig{pp_to_ww:14TeV:weights}, we visualize the utility of mass EWOCs with non-unity energy weights to mitigate the effects of UE, as discussed in \Sec{EWOCs:IRC}.
Figs.~\ref{fig:m_ewoc:rsub_3:weight_2} and \ref{fig:m_ewoc:rsub_3:weight_3} show plots of the mass EWOC with \(\rsub = 0.3\) with energy weights \(n = m = 2\) and \(n = m = 3\), respectively.
Table \ref{tab:energyweights} tabulates the shifts in the peaks of the mass EWOC for these energy weights due to both momentum smearing and non-perturbative effects, and compares these shifts to the analogous shifts in the mass EWOC with unity energy weights.
We see that the use of higher energy weights generates results that are slightly more resistant to both to our rough model of detector effects and the non-perturbative effects of hadronization and the UE.

We conclude that the mass EWOC is a robust phenomenological tool for the extraction of mass scales in particle collisions.
The use of subjets leads to results which are resilient to non-perturbative and experimental effects, and are quickly and intuitively interpretable as mass scales.
Now, we turn towards mass EWOCs in the cleaner laboratory of electron-positron collisions in order to gain a deeper understanding of their analytic structure.

% ==============================================
\subsection{Perturbative Results}
% ==============================================
\label{sec:analytic}

% -----------------------------------
% LO and All-Orders Visual Representations
% -----------------------------------
\begin{figure}
    \centering
    \hspace{-1.0cm}
    \subfloat[]{
        \scalebox{0.9}{
            \input{figures/calculation/lo_phase_space.tikz}
        }
        \label{fig:ee2hadrons:phase_space}
    }
    \hspace{0.0cm}
    \subfloat[]{
    \begin{tikzpicture}
        \node[scale=1.3] (allorders) {
            \begin{tikzpicture}
\begin{scope}[scale=0.5]
    % Jet visualization
    % Variables
    \def\x{2.00}
    \def\y{3.335}
    \def\R{\x+0.005}
    \def\yc{\y+0.04}
    \def\e{0.5}
    
    % Jet
    \begin{scope}[scale=1.5,rotate=-90]
        \clip (0.0,0.0) ++(0:0.2) arc (0:180:0.2) -- (-3,4.0) -- (3,4.0) -- cycle;
        \shade[right color=white,left color=black,opacity=0.1]
          (-\x,\yc) -- (-\x,\yc) arc (180:360:{\R} and \e) -- (\x,\yc) -- (0,0) -- cycle;
        \draw[fill=black,opacity=0.05]
          (0,\yc) circle ({\R} and \e);
        \draw
          (-\x,\y) -- (0,0) -- (\x,\y);
        \draw
          (0,\yc) circle ({\R} and \e);
    \end{scope}

    \begin{feynman}
        \vertex (li);
        \vertex [blob, scale=0.45, right=0.8 cm of li] (a) {};
        \vertex [dot, scale=0.3, right=0.75 cm of a] (d) {};
        \vertex [blob, scale=0.25, above right=.25cm and 0.5 cm of d] (e) {};
        \vertex [blob, scale=0.25, below right=.25cm and 0.5 cm of d] (f) {};
        \vertex [above right=.1cm and .95cm of e] (g) {};
        \vertex [below right=.1cm and .95cm of f] (h) {};
        \diagram* {
            (li) -- (a),
            (a)  --  (d),
            (a)  -- [gluon]  (d),
            (d)  -- (e),
            (d)  -- [gluon] (e),
            (d)  -- (f),
            (d)  -- [gluon] (f),
            (e) -- (g),
            (e) -- [gluon] (g),
            (f) -- (h),
            (f) -- [gluon] (h),
        };
    \end{feynman}

    % Subjet visualization
    \begin{scope}[xshift=20pt,yshift=-3.1pt]
        % Inclusive over states other than final splitting
        \def\x{1.2}
        \def\y{3.4}
        \def\R{\x+0.005}
        \def\yc{\y+0.04}
        \def\e{0.4}
        \begin{scope}[xshift=1.32cm,yshift=.12cm]
        \begin{scope}[scale=0.35,rotate=-34]
        \clip (0.0,0.0) ++(0:0.8) arc (0:180:0.8) -- (-3,4) -- (3,4) -- cycle;
        \shade[right color=white,left color=inclusivecolor!70!black,opacity=0.7]
          (-\x,\yc) -- (-\x,\yc) arc (180:360:{\R} and \e) -- (\x,\yc) -- (0,0) -- cycle;
        \draw[fill=inclusivecolor!70!black,opacity=0.15]
          (0,\yc) circle ({\R} and \e);
        \draw
          (-\x,\y) -- (0,0) -- (\x,\y);
        \draw
          (0,\yc) circle ({\R} and \e);
        \end{scope}
        \end{scope}

        % Radiation post-splitting
        \def\x{0.5}
        \def\y{3.0}
        \def\R{\x+0.005}
        \def\yc{\y+0.04}
        \def\e{0.2}
       
        \begin{scope}[xshift=3.63cm,yshift=11pt]
        \begin{scope}[scale=0.4,rotate=-25]
        \clip (0.0,0.0) ++(0:1) arc (0:180:1) -- (-2,4) -- (2,4) -- cycle;
        \shade[right color=white,left color=inclusivecolor!70!black,opacity=0.7]
          (-\x,\yc) -- (-\x,\yc) arc (180:360:{\R} and \e) -- (\x,\yc) -- (0,0) -- cycle;
        \draw[fill=inclusivecolor!70!black,opacity=0.15]
          (0,\yc) circle ({\R} and \e);
        \draw
          (-\x,\y) -- (0,0) -- (\x,\y);
        \draw
          (0,\yc) circle ({\R} and \e);
        \end{scope}
        \end{scope} 
        
        \begin{scope}[xshift=3.63cm,yshift=-5.0pt]
        \begin{scope}[scale=0.4,rotate=-155]
            \clip (0.0,0.0) ++(0:1) arc (0:180:1) -- (-2,4) -- (2,4) -- cycle;
            \shade[right color=white,left color=inclusivecolor!70!black,opacity=0.7]
              (-\x,\yc) -- (-\x,\yc) arc (180:360:{\R} and \e) -- (\x,\yc) -- (0,0) -- cycle;
            \draw[fill=inclusivecolor!70!black,opacity=0.15]
              (0,\yc) circle ({\R} and \e);
            \draw
              (-\x,\y) -- (0,0) -- (\x,\y);
            \draw
              (0,\yc) circle ({\R} and \e);
        \end{scope}
        \end{scope}

        % Subjets Emerging from Splitting
        \def\x{0.9}
        \def\y{3.0}
        \def\R{\x+0.005}
        \def\yc{\y+0.04}
        \def\e{0.3}
       
        % \begin{scope}[xshift=4.5cm,yshift=20.0pt]
        \begin{scope}[xshift=4.1cm,yshift=18.0pt]
        \begin{scope}[scale=0.4,rotate=-84]
            % \clip (0.0,0.0) ++(0:1) arc (0:180:1) -- (-2,4) -- (2,4) -- cycle;
            \shade[right color=white,left color=green,opacity=0.3]
              (-\x,\yc) -- (-\x,\yc) arc (180:360:{\R} and \e) -- (\x,\yc) -- (0,0) -- cycle;
            \draw[fill=green,opacity=0.2]
              (0,\yc) circle ({\R} and \e);
            \draw[green!30!black]
              (-\x,\y) -- (0,0) -- (\x,\y);
            \draw[green!30!black]
              (0,\yc) circle ({\R} and \e);
        \end{scope}
        \end{scope}
        
        % \begin{scope}[xshift=4.5cm,yshift=-14.0pt]
        \begin{scope}[xshift=4.1cm,yshift=-12.0pt]
        \begin{scope}[scale=0.4,rotate=-97]
            % \clip (0.0,0.0) ++(0:1) arc (0:180:1) -- (-2,4) -- (2,4) -- cycle;
            \shade[right color=white,left color=green,opacity=0.3]
              (-\x,\yc) -- (-\x,\yc) arc (180:360:{\R} and \e) -- (\x,\yc) -- (0,0) -- cycle;
            \draw[fill=green,opacity=0.2]
              (0,\yc) circle ({\R} and \e);
            \draw[green!30!black]
              (-\x,\y) -- (0,0) -- (\x,\y);
            \draw[green!30!black]
              (0,\yc) circle ({\R} and \e);
        \end{scope}
        \end{scope}

    \end{scope}
\end{scope}

\end{tikzpicture}
        };
        
        \node[scale=1.4, left=0.0cm of allorders] (math) {
            \(
            \frac{\dd \Sigma^\text{LL}_\mc{O}}{\dd \chi}
            \,\,=\,\,
            \)
        };
        \node[scale=0.8, above right=-13pt and -26.5pt of math] {coll.};
        \node[scale=0.8, below right=-13pt and -27.5pt of math] {limit};
        
        \node[scale=1.35, below right=1.3 and -0.5cm of math] (text) {
            \textcolor{inclusivecolor}{where}
        };
        \node[scale=0.95, below left=-0.3cm and -4.7cm of allorders] (fragmentation) {
            % Inclusive diagram
\raisebox{6pt}{
\begin{tikzpicture}
    \begin{feynman}
        \vertex (li);
        \vertex [blob, scale=0.45, right=0.7 cm of li] (a) {};
        \vertex [scale=0.3, right=0.8 cm of a] (d) {};
        \diagram* {
            (li) -- (a),
            (a)  --  (d),
            (li)  -- [gluon]  (a),
            (a)  -- [gluon]  (d),
        };
    \end{feynman}

    % Subjet visualization
    \begin{scope}[xshift=-27pt,yshift=-5pt]
        % Inclusive over states other than final splitting
        \def\x{1.2}
        \def\y{3.4}
        \def\R{\x+0.005}
        \def\yc{\y+0.04}
        \def\e{0.4}
        \begin{scope}[xshift=1.8cm,yshift=.12cm]
        \begin{scope}[scale=0.26,rotate=-30]
        \clip (0.0,0.0) ++(0:0.8) arc (0:180:0.8) -- (-3,4) -- (3,4) -- cycle;
        \shade[right color=white,left color=inclusivecolor!70!black,opacity=0.7]
          (-\x,\yc) -- (-\x,\yc) arc (180:360:{\R} and \e) -- (\x,\yc) -- (0,0) -- cycle;
        \draw[fill=inclusivecolor!70!black,opacity=0.15]
          (0,\yc) circle ({\R} and \e);
        \draw
          (-\x,\y) -- (0,0) -- (\x,\y);
        \draw
          (0,\yc) circle ({\R} and \e);
        \end{scope}
        \end{scope}
    \end{scope}
\end{tikzpicture}
}

\hspace{-18pt}
\raisebox{6pt}{\scalebox{1.3}{\(=\)}}
\hspace{-5pt}

% Sum with description
% \raisebox{7pt}{ \scalebox{1.6}{\(\sum\)} }

% \hspace{-35pt}
% \raisebox{23pt}{\scalebox{0.75}{\textcolor{inclusivecolor}{inclusive}}}

% \hspace{-40pt}
% \raisebox{-6pt}{\scalebox{0.75}{\textcolor{inclusivecolor!80!black}{radiation}}}

% Explicit splittings
% \begin{tikzpicture}[very thick, inclusivecolor!120!black!80!white]
    % \begin{feynman}
    %     \vertex (li);
    %     \vertex [right=0.5 cm of li] (b);
    %     \vertex [above right=.48cm and .2cm of b] (ab);
    %     \vertex [right=0.3 cm of b] (c);
    %     \vertex [below right=.4cm and .3cm of c] (ac);
    %     \vertex [right=0.1 cm of c] (ct);
    %     \vertex [right=.5 cm of ct] (ctt);
    %     \vertex [right=.3 cm of ctt] (d);
    %     \vertex [above right=.3cm and .4cm of ctt] (ad);
    %     \vertex [right=0.5 cm of d] (lf);
    %     \diagram* {
    %         % Horizontal lines
    %         (li) -- [thin, color=black] (b),
    %         (li) -- [thin, gluon, color=black] (b),
    %         (b) -- (c) -- (ct) -- [dotted] (ctt) -- (d),
    %         (b)  -- [gluon] (c) -- [gluon] (ct), (ctt) -- [gluon] (d),
    %         (d) -- [thin, color=black] (lf), (d) -- [thin, gluon, color=black] (lf),
    %         % Outgoing Legs
    %         (b)  -- (ab), (b)  -- [gluon] (ab),
    %         (c)  -- (ac), (c)  -- [gluon] (ac),
    %         (ctt)-- (ad), (ctt)-- [gluon] (ad)
    %     };
    % \end{feynman}
% \end{tikzpicture}

\raisebox{8pt}{
\begin{tikzpicture}[very thick, inclusivecolor!120!black!80!white]
    \begin{feynman}
        \vertex (li);
        \vertex [dot, scale=0.3, right=0.5 cm of li, color=black] (b) {};
        \vertex [above right=.48cm and .2cm of b] (ab);
        \vertex [right=0.55 cm of b] (f);
        \diagram* {
            % Horizontal lines
            (li) -- [thin, color=black] (b),
            (li) -- [thin, gluon, color=black] (b),
            (b) -- [thin, color=black] (f),
            (b) -- [thin, gluon, color=black] (f),
            % Outgoing Legs
            (b)  -- (ab),
            (b)  -- [gluon] (ab),
        };
        % Draw vertices on top
        \draw[dot, color=black] (b) circle (0.3mm);
    \end{feynman}
\end{tikzpicture}
}

\hspace{-10pt}
\raisebox{7pt}{ \scalebox{1.3}{\(+\)} }
\hspace{-10pt}
    
\raisebox{-2pt}{
\begin{tikzpicture}[very thick, inclusivecolor!120!black!80!white]
    \begin{feynman}
        \vertex (li);
        \vertex [dot, scale=0.3, right=0.5 cm of li, color=black] (b) {};
        \vertex [above right=.48cm and .2cm of b] (ab);
        \vertex [dot, scale=0.3, right=0.3 cm of b, color=black] (c) {};
        \vertex [below right=.4cm and .3cm of c] (ac);
        \vertex [right=0.55 cm of c] (lf);
        \diagram* {
            % Horizontal lines
            (li) -- [thin, color=black] (b),
            (li) -- [thin, gluon, color=black] (b),
            (b) -- (c),
            (b)  -- [gluon] (c),
            (c) -- [thin, color=black] (lf), (c) -- [thin, gluon, color=black] (lf),
            % Outgoing Legs
            (b)  -- (ab), (b)  -- [gluon] (ab),
            (c)  -- (ac), (c)  -- [gluon] (ac),
        };
        % Draw vertices on top
        \draw[dot, color=black] (b) circle (0.3mm);
        \draw[dot, color=black] (c) circle (0.3mm);
    \end{feynman}
\end{tikzpicture}
}
    
\hspace{-10pt}
\raisebox{7pt}{ \scalebox{1.3}{\(+\,\cdots\)} }
        };
        \node[below=0.5cm of fragmentation] (bottom) {};
    \end{tikzpicture}    
    \label{fig:ee2hadrons:allorders}
    }
    \caption{
        Ingredients in the calculation of EWOCs \hyperref[fig:ee2hadrons:phase_space]{(a)} at leading order (LO) and \hyperref[fig:ee2hadrons:allorders]{(b)} at all orders.
        \hyperref[fig:ee2hadrons:phase_space]{(a)} The two-parton phase space can be divided into three regions, in which the two partons are contained either within the same subjet (orange), different subjets (green), or different jets (red).
        \hyperref[fig:ee2hadrons:allorders]{(b)} A diagrammatic visualization of a leading-logarithmic computation of EWOCs in the collinear limit, by convolving QCD splitting functions with several instances of parton-to-parton fragmentation, discussed in greater detail in the text.
        Combined solid/coiled lines indicate a line of arbitrary partonic flavor, blue cones indicate the inclusive emission of a parton, and the three-point vertex denotes a leading-order QCD splitting function.
    }
    \label{fig:EWOCs:ee2hadrons:visualization}
\end{figure}
% -----------------------------------

% -----------------------------------
% LO and Pythia Plots 
% -----------------------------------
\begin{figure}
    \hspace{-20pt}
    \subfloat[]{
        \includegraphics[width=.53\textwidth]{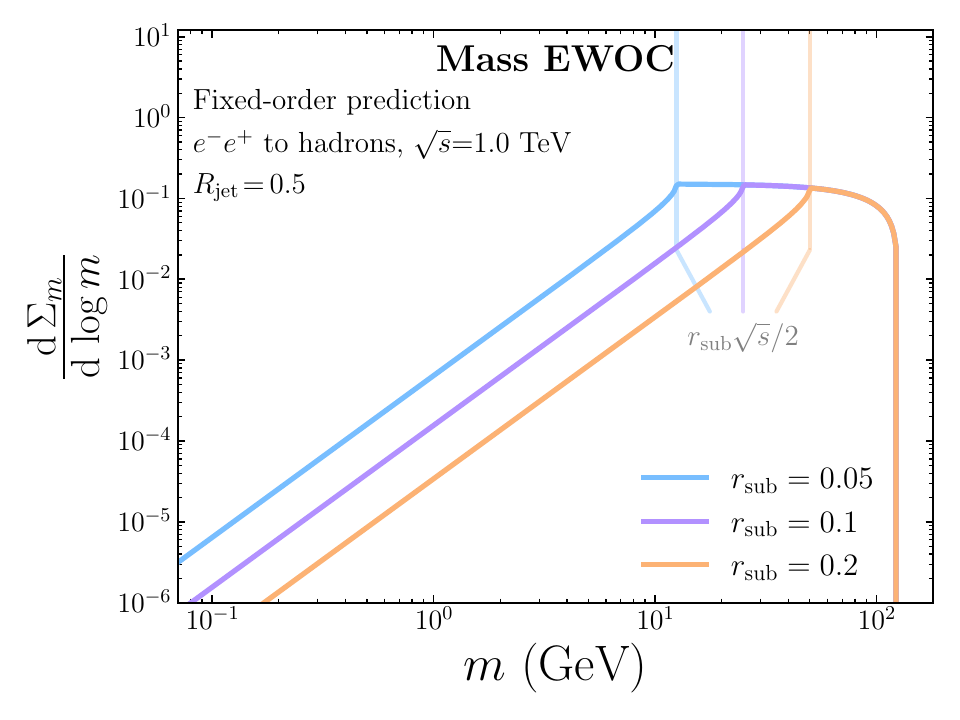}
        \label{fig:ee2hadrons:lo}
    }
    \subfloat[]{
        \includegraphics[width=.53\textwidth]{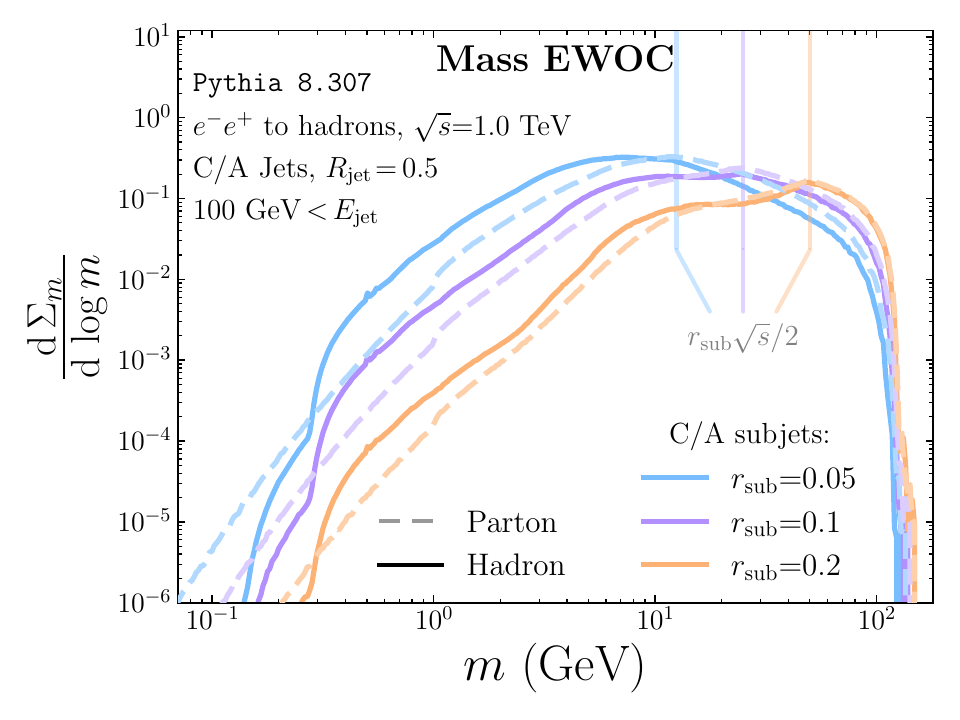}
        \label{fig:ee2hadrons:pythia}
    }
    \caption{
        Mass EWOCs for \(e^+ e^- \to \,\)hadrons \hyperref[fig:ee2hadrons:lo]{(a)} at LO and \hyperref[fig:ee2hadrons:pythia]{(b)} in \pythia{}.
    }
    \label{fig:EWOCs:ee2hadrons}
\end{figure}
% -----------------------------------

Now that we have presented the story of EWOCs in a realistic phenomenological example, we compare the features of mass EWOCs in light-quark-initiated jets obtained both perturbatively at leading order (LO) in $\alpha_s$ and with \pythia{} in the standard laboratory of \(e^+ e^-\to\,\)hadrons.
We expect that more detailed, leading-logarithmic results will depend on numerical analyses involving parton-to-parton fragmentation (e.g.~the semi-inclusive jet function of \Reff{Kang:2016mcy} or the microjet fragmentation functions of refs.~\cite{Dasgupta:2014yra,Dasgupta:2016bnd}), which we briefly outline at the end of this section.

The analysis we explore in this section involves jets consisting of two partons.
The corresponding phase space may be separated into three regions as depicted in \Fig{ee2hadrons:phase_space}:
\begin{itemize}
    \item 
    the region with \(\theta < \rsub\) (the leftmost, orange region of \Fig{ee2hadrons:phase_space}), where the two partons are grouped into the same subjet;

    \item
    the region with \(\rsub < \theta < \Rjet\) (the middle, green region), where the two partons are within the same jet but different subjets;

    \item 
    the region with \(\Rjet < \theta\) (the rightmost, red region), where the two partons are grouped into different jets.
\end{itemize}
We will use a subjet recombination scheme that assigns zero mass to subjets -- in our study of \(e^+ e^-\) collisions, the winner-take-all (WTA) \(|\vec{p}|\) recombination scheme is a natural choice -- to avoid complications due to subjet mass spectra.
The non-trivial contribution to the EWOC then comes from the region of phase space when the two partons are separated by an angle \(\rsub < \theta < \Rjet\) and are constructed as distinct subjets within the same jet.
The other two regions, in which the partons are reconstructed as distinct jets or as within the same subjet, contribute via ``contact terms'':
delta-functions, due to virtual corrections as well as 2-particle subjet self-correlations, whose combined coefficient is fixed by the fact that the EWOC integrates to one.
For example, in the case of mass EWOC, $\dd \Sigma_m / \dd m$, the contact-term regions produce zero-mass contributions of the form \(\le(1 + \mathcal{O}(\alpha_s)\ri)\delta(m)\).
We note, however, that if we used a subjet recombination scheme which produced massive subjets, the contact terms would no longer be delta functions at \(m = 0\) and the analysis of the resulting mass EWOC would be more difficult.

For (massless) quark-initiated jets, we find that the LO mass EWOC for \(m > 0\) is given by
\begin{align}
    \frac{\dd\Sigma^\text{LO}_m}{\dd m}(m > 0)
    &=
    \frac{3 \alpha_s \, C_F}{8\pi\, m}
    \begin{cases}
        % \le(1 - \frac{8}{3}\frac{m^2}{s \, \Rjet^2}\ri)
        % \sqrt{1 - \frac{16 m^2}{s \, \Rjet^2}} 
        V(m, \, s\,\Rjet^2)
        \,,
        \quad\qquad\qquad\qquad
        \sqrt{s}\,\rsub/4
        \, <
        m
        <
        \sqrt{s}\,\Rjet/4
        \,,
        \\
        % \le(1 - \frac{8}{3}\frac{m^2}{s \, \Rjet^2}\ri)
        % \sqrt{1 - \frac{16 m^2}{s \, \Rjet^2}} 
        V(m, \, s\,\Rjet^2)
        -
        V(m, \, s\,\rsub^2)
        % \le(1 - \frac{8}{3}\frac{m^2}{s \, \rsub^2}\ri)
        % \sqrt{1 - \frac{16 m^2}{s  \, \rsub^2}} 
        \,,
        \quad
        0 < m \, < \sqrt{s}\,\rsub/4
        \,,
    \end{cases}
\end{align}
where \(
    V(m, Q^2)
    \equiv 
    \le(1 - 8 m^2 / \le(3 Q^2\ri)\ri)
    \sqrt{1 - 16 m^2 / Q^2} 
\) and $\sqrt{s}$ is the center-of-mass energy of the collision.
\(\dd \Sigma^\text{LO}_m / \dd m\) is also shown in \Fig{ee2hadrons:lo}.\footnote{
    Since the mass EWOC is guaranteed to integrate to one, the correctly normalized mass EWOC (including contributions at \(m=0\)) can always be written as \(
        \dd\Sigma_m / \dd m
        =
        \delta(m)
        +
        \le[ \,
        \dd\Sigma_m / \dd m (m > 0)
        \, \ri]_+
        \,
    \), where \(\le[ \, f(m) \, \ri]_+\) indicates a plus-regularization of the function \(f\) which integrates to zero.
}

The mass EWOC for quark-initiated jets obtained via \pythia{} simulations of \(e^+e^-\to\,\)hadrons is shown in \Fig{ee2hadrons:pythia}.
The fixed-order mass EWOC has a kink at the mass scale \(\sqrt{s}\,\,\rsub/2\),
the largest mass scale at which the two partons of the fixed-order calculation can be grouped into a single subjet of radius \rsub{}, which separates the domain of the mass EWOC into two distinct regions.
We expect that this kink is smoothed out by the effects of multiple emissions,\footnote{
    This expectation is supported by our rough initial studies of the analytic behavior of EWOCs at all orders, but we delegate a more rigorous justification to future work;
    similar results involving all-orders smoothing of kinks found in fixed-order distributions of jet substructure observables can be found in e.g.~\Reff{Benkendorfer:2021unv}.
} and indeed the mass EWOCs obtained with \pythia{} appear smooth at \(m = \sqrt{s}\,\,\rsub/2\).
Nonetheless, the shapes of the fixed-order and \pythia{} mass EWOCs are roughly the same, and the power laws that roughly govern the fixed-order mass EWOC are also present in the mass EWOC obtained with \pythia{}.

Having discussed the LO EWOC, we now briefly discuss the ingredients involved in a leading-logarithmic (LL) computation, which requires a more detailed numerical analysis of partonic fragmentation including the possibility of arbitrarily many partonic emissions.
As visualized diagramatically in fig.~\ref{fig:ee2hadrons:allorders}, at LL accuracy and in the regime $\sqrt{s} \sim \sqrt{s} \Rjet \gg m \gg \sqrt{s} \rsub \gg \Lambda_{\rm QCD}$, this entails the computation of intricate convolution integrals of the form
\begin{align}
    \label{eq:allorder_ewoc_convolution}
    \frac{\dd \Sigma^\text{LL}_m}{\dd m}
    =&
    \sum_j 
    \int_0^1 \dd z_1\,    
    z_1^2\, F_{j \leftarrow \text{q}}
    (z_1;\,\theta\leftarrow \Rjet)
    \sum_{k,\ell}
    \frac{\alpha_s}{\pi}
    \int_0^1 \dd z'\,
    z^{\prime}(1-z^\prime)
    \int_{\rsub}^{\Rjet}
    \frac{\dd \theta}{\theta}
    \,
    P_{k \ell \leftarrow j}(z')
    \notag \\ & \times
    \sum_c
    \int_0^1 \dd z_2\, z_2\,
    F_{c \leftarrow k}
    (z_2;\,\rsub \leftarrow \theta)
    \sum_d
    \int_0^1 \dd z_3\, z_3\,
    F_{d \leftarrow \ell}
    (z_3;\,\rsub \leftarrow \theta)
    \notag \\ & \quad \times
    \delta\biggl(
        m
        \,\,
        -
        \,\,
        \frac{\sqrt{s}}{2}
        \,
        z_1 
        \,
        \sqrt{
        z^\prime
        (1-z^\prime)
        }
        \,
        \theta
        \,
        \sqrt{z_2 \, z_3}
    \biggr)    
    \,
    ,
\end{align}
which adds up the contributions to the mass EWOC from all possible branching histories of a quark-initiated jet (in the collinear limit) and all pairs of final state subjets.
In \Eq{allorder_ewoc_convolution}, \(P_{k \ell \leftarrow j}(z_k)\) is the leading order Dokshitzer-Gribov-Lipatov-Altarelli-Parisi (DGLAP) splitting function \cite{Gribov:1972ri,Dokshitzer:1977sg,Altarelli:1977zs} which encodes the pseudo-probability for a parton of type \(i\) to split into a pair of partons of types \(j\) and \(k\) with the energy fraction \(z_k = 1 - z_\ell = E_k / E_j\). 
In QCD, \(\ell\) is fixed by \(j\) and \(k\), and we can write \(P_{k \ell \leftarrow j}(z) = p_{k \leftarrow j}(z)\).%
\footnote{
    For example, \(P_{qg\leftarrow q}\) is non-zero, but \(P_{gg \leftarrow q}\) is zero.
}
The \(F_{j \leftarrow i}(z;\,\theta_j \leftarrow \theta_i)\) are parton-to-parton fragmentation functions which obey the DGLAP evolution equations and encode the probability density that a parton of type \(i\), probed at an angular resolution \(\theta_i\), contains a parton of type \(j\) with energy fraction \(z\) when probed at an angular resolution of \(\theta_j < \theta_i\).
The use of parton-to-parton fragmentation functions resums both observable-related logarithms (such as \(\ln m^2 /s\)) and (sub)jet-radius logarithms (\(\ln\Rjet\) and \(\ln\rsub\)).
The resummed form of the parton-to-parton fragmentation functions required for the computation of \Eq{allorder_ewoc_convolution} is, to our knowledge, obtained only through numeric solutions to the DGLAP equation in the existing literature.%
\footnote{
    Beyond LL accuracy, the resummation of parton-to-parton fragmentation functions is not  described by standard DGLAP,
    see e.g.~refs.~\cite{vanBeekveld:2024jnx,vanBeekveld:2024qxs,Lee:2024icn,Lee:2024tzc}.
}

In the case of the EEC, the argument of the delta function in  \Eq{allorder_ewoc_convolution} depends only on the angle \(\theta\), and not on any of the energy fractions.
In this case, the momentum-fraction integrals decouple and become \textit{Mellin moments}
\(
    \hat{F}_{j \leftarrow i}(\kappa)
    \equiv
    \int_0^1 \dd z \, z^{\kappa-1}\, F_{j \leftarrow i}(z) 
\)
of the parton-to-parton fragmentation functions, whose evolution is determined by the Mellin-space form of the DGLAP evolution equations,
\begin{align}
    \frac{\dd}{\dd\log\, r} 
    \hat{F}_{j\leftarrow i}(\kappa;\,
    r\leftarrow R)
    &=
    \sum_k
    \frac{\alpha_s}{\pi}
    \,\,
    \hat{F}_{j\leftarrow k}(\kappa;\,r\leftarrow R)
    \,\,
    \hat{p}_{k\leftarrow i}(\kappa)
    \,
    ,
\end{align}
This leads quickly to the leading-logarithmic form \(
    \hat{F}_{j\leftarrow i}(\kappa ;\,
    r\leftarrow R)
    \,
    =
    \,
    \exp\le[
        \alpha_s
        \,\,
        \hat{p}(\kappa)
        \,
        \log\le(r/R\ri)
        /
        \pi
    \ri]_{j \leftarrow i}
\) for the parton-to-parton fragmentation functions\footnote{
See, for example, table III of \Reff{Konishi:1978ks} for expressions for the \(\hat P_{j\ell\leftarrow i}(\kappa)\).
}, and reproduces the leading-logarithmic EEC obtained via jet calculus \cite{Konishi:1978dg,Konishi:1978yx,Konishi:1978ks}.
However, for more generic EWOCs, the momentum-fraction integrals do not decouple, and numerical integration of \Eq{allorder_ewoc_convolution} will be required to obtain more precise, leading-logarithmic results.

%%%%%%%%%%%%%%%%%%%%%%%%
% Discussion and Conclusions:
%%%%%%%%%%%%%%%%%%%%%%%%
\section{Discussion and Conclusions}
\label{sec:Conclusions}

% ============================
% In this work, we
% ============================
In this work, we introduced Energy-Weighted Observable Correlations (EWOCs) as a flexible new tool for studying the patterns of radiation within hadronic jets produced in particle collisions, generalizing the more familiar Energy-Energy Correlator (EEC).
We examined the formal features of EWOCs, their utility in the extraction of mass scales in particle collisions, and their analytic structure.

We began by motivating and defining EWOCs as generalizations of the EEC which can be chosen to probe a variety of different multi-particle correlations, such as the dominant mass scales or formation times involved in jet production in a particular jet sample.
We highlighted that collinear safety requires generic EWOCs to be equipped with an additional collinear cutoff, such as a subjet radius.
Consequently, the subjet EWOCs on which we focus in this work are more robust to non-perturbative effects such as hadronization.
We also found that EWOCs, unlike the particle-level EEC, remain collinear-safe even when using non-unity energy weights, allowing for additional suppression of obfuscating soft physics such as hadronization and the underlying event (UE).

We then restricted ourselves to the study of the mass EWOC.
We showed that the mass EWOC provided an estimate of the \(W\) boson mass which was even more robust to the effects of hadronization and UE than the analogous estimate using the EEC (and comparable to the estimate of \(m_W\) using groomed jet masses).
We also showed that EWOCs with higher energy weights were more robust to hadronization, UE, and a rough model of smearing due to experimental detectors.
We then presented the mass EWOC at leading order in \(e^+ e^- \to\,\)hadrons -- the simplest testbed of perturbative QCD.
We conclude that EWOCs provide new perspectives into physical correlations within jets which are robust to soft radiation, and are flexible probes of physics in a variety of contexts due to the user-defined observable and subjet radius.

There are several straightforward avenues for future work:
Extending the analysis of \Sec{mass} to real collision data and including the effects of QCD backgrounds would provide a more realistic test of the utility of the mass EWOC in extracting the \(W\) mass and beyond.
The most precise existing measurements of the \(W\) boson mass involve leptonic decays of the \(W\) boson and require a detailed understanding of recoil and missing energy \cite{CMS:2024nau}.
The use of the mass EWOC allows us instead to study the mass of the \(W\) boson through its hadronic decays, and without sensitivity to recoil.%
\footnote{
    We thank Christoph Paus for emphasizing this feature to us.
}
The study of more complicated observables with EWOCs in more variegated contexts is also a promising approach towards extending the utility of EWOCs;
for example, the study of formation-time EWOCs in heavy-ion collisions may elucidate time scales associated with the interactions of high-energy jets and the quark-gluon plasma. 
More broadly, the use of energy flow polynomials \cite{Komiske:2017aww} as EWOC observables could provide a powerful new tool for studying a variety of correlations within jets.

A particularly simple and phenomenologically interesting extension of the analysis presented in this work is the study of EWOCs to three-point observables;
the three-point mass EWOC, for example, could be used to extract the top quark mass from LHC data.
We also note that EWOCs may have applications in searches for physics beyond the Standard Model (BSM).
In particular, for models which predict the production of SM jets from the decay of BSM particles, the substructure of the SM jets may be used to probe the properties of the BSM particles.
It is possible that our proof of concept for the mass EWOC as a tool for extracting the \(W\)-boson mass from LHC collisions in this work may be extended to the extraction of mass scales involved in BSM decays.

This paper serves as an invitation to further explore the landscape of EWOCs and their applications.
We expect that EWOCs and their generalizations can contribute to the bridge between experimental and theoretical particle physics, providing new insights into nature from particle collisions.

%%%%%%%%%%%%%%%%%%%%%
\acknowledgments
%%%%%%%%%%%%%%%%%%%%%
We thank Ian Moult for collaboration in early stages of this project, Jesse Thaler for useful insight and feedback on the manuscript, and Massimiliano Procura for helpful conversations.
S.A.F.\ is supported by the U.S. DOE Office of High Energy Physics under grant number DE-SC0012567.

\bibliography{ewoc_paper}

\end{document}